\documentclass[a4paper, 12pt]{article}
\pdfoutput=1

\usepackage{latexsym,amsmath,amsfonts,amssymb}
\usepackage{tikz}
\usetikzlibrary{decorations.pathmorphing,cd,decorations.markings}
\usepackage[latin1]{inputenc}
\usepackage[american]{babel}
\usepackage{graphicx}
\usepackage{bbm}
\usepackage{cite}
\usepackage[colorlinks=true, citecolor=blue, linkcolor=blue, linktocpage=true]{hyperref}
\usepackage{longtable}
\usepackage{tikz,tkz-euclide,tikz-cd}
\usepackage{circuitikz}
\usetikzlibrary{calc}
\usetikzlibrary{decorations.pathmorphing, decorations.markings}
\usetikzlibrary{patterns}
\usepackage{soul}
\usepackage{comment}

\renewcommand{\baselinestretch}{1.2}
\newcommand{\ds}{\displaystyle}

\newcommand{\fP}{\mathfrak{P}}

\newcommand{\eg}{\textit{e.g.}}

\newcommand{\ie}{\textit{i.e.}}

\numberwithin{equation}{section}

\newcommand{\mat}[1]{\begin{pmatrix} #1 \end{pmatrix}}

\newcommand{\be}{\begin{equation}} \newcommand{\ee}{\end{equation}}
\newcommand{\bea}{\begin{equation} \begin{aligned}} \newcommand{\eea}{\end{aligned} \end{equation}}

\newcommand{\cA}{\mathcal{A}}
\newcommand{\cB}{\mathcal{B}}
\newcommand{\cC}{\mathcal{C}}
\newcommand{\cD}{\mathcal{D}}
\newcommand{\cE}{\mathcal{E}}

\newcommand{\cH}{\mathcal{H}}

\newcommand{\cJ}{\mathcal{J}}
\newcommand{\cK}{\mathcal{K}}
\newcommand{\cL}{\mathcal{L}}
\newcommand{\cM}{\mathcal{M}}
\newcommand{\cN}{\mathcal{N}}

\newcommand{\cP}{\mathcal{P}}

\newcommand{\cR}{\mathcal{R}}
\newcommand{\cS}{\mathcal{S}}
\newcommand{\cT}{\mathcal{T}}
\newcommand{\cU}{\mathcal{U}}

\newcommand{\cW}{\mathcal{W}}

\newcommand{\bR}{\mathbb{R}}

\newcommand{\bZ}{\mathbb{Z}}
\newcommand{\fD}{\mathfrak{D}}

\newcommand{\sT}{\mathsf{T}}

\newcommand{\unit}{\mathbbm{1}}
\newcommand{\tor}{\text{Tor}}
\newcommand{\MCG}{\text{MCG}}

\def\SU{\mathrm{SU}}
\def\PSU{\mathrm{PSU}}
\def\SL{\mathrm{SL}}
\def\GL{\mathrm{GL}}
\def\Sp{\mathrm{Sp}}

\def\repa{\raise4pt\hbox{$\square$}\mkern-14mu\raise-4pt\hbox{$\square$}}
\def\repab{\overline{\raise4pt\hbox{$\square$}\mkern-14mu\raise-4pt\hbox{$\square$}\mkern-1mu}}

\newcommand{\smallwedge}{\mathrel{\text{\raisebox{0.25ex}{\scalebox{0.75}{$\wedge$}}}}}
\newcommand{\smallvee}{\mathrel{\text{\raisebox{0.25ex}{\scalebox{0.75}{$\vee$}}}}}

\newcommand{\inv}{^{\raisebox{.2ex}{$\scriptscriptstyle-1$}}}

\begin{document}
\thispagestyle{empty}
\fontsize{12pt}{20pt}

\vspace{13mm}  
\begin{center}
	{\huge ``Zoology" of non-invertible duality defects: \\[.5em] the view from class $\cS$}
	\\[13mm]
    	{\large Andrea Antinucci$^{a, \, b} \,$, Christian Copetti$^{a, \, b} \,$, Giovanni Galati$^{a, \, b} \,$, \newline Giovanni Rizi$^{a, \, b} \,$} 
	
	\bigskip
	{\it
		$^a$ SISSA, Via Bonomea 265, 34136 Trieste, Italy \\[.2em]
		$^b$ INFN, Sezione di Trieste, Via Valerio 2, 34127 Trieste, Italy \\[.2em]
	}
\end{center}

\begin{abstract}
\noindent We study generalizations of the non-invertible duality defects present in $\cN = 4$ $\SU(N)$ SYM by studying theories with larger duality groups. We focus on 4d $\cN=2$ theories of class $\cS$ obtained by the dimensional reduction of the 6d $\cN=(2,0)$ theory of $A_{N-1}$ type on a Riemann surface $\Sigma_g$ without punctures.
We discuss their non-invertible duality symmetries and provide two ways to compute their fusion algebra: either using discrete topological manipulations or a 5d TQFT description. We also introduce the concept of ``rank'' of a non-invertible duality symmetry and show how it can be used to (almost) completely fix the fusion algebra with little computational effort.
\end{abstract}

\newpage

{\renewcommand{\baselinestretch}{.88} \parskip=0pt
\setcounter{tocdepth}{2}
\tableofcontents}

\pagebreak

\section{Introduction}\label{introduction}
Generalizations of the concept of symmetry have attracted a great deal of attention since the seminal work \cite{Gaiotto:2014kfa}, in which symmetries are defined as the set of topological operators of a given quantum field theory. Following this line of thought many recent works have described how to construct symmetry defects which do not follow a group law in $d>2$ \cite{ Roumpedakis:2022aik, Bhardwaj:2022yxj, Arias-Tamargo:2022nlf, Choi:2022zal, Kaidi:2022uux, Choi:2022jqy, Cordova:2022ieu, Antinucci:2022eat, Bashmakov:2022jtl, Damia:2022rxw, Damia:2022bcd, Choi:2022rfe, Bhardwaj:2022lsg, Lin:2022xod, Bartsch:2022mpm, Apruzzi:2022rei, GarciaEtxebarria:2022vzq, Heckman:2022muc, Niro:2022ctq, Antinucci:2022vyk, Chen:2022cyw, Cordova:2022fhg, GarciaEtxebarria:2022jky, Choi:2022fgx, Yokokura:2022alv, Bhardwaj:2022kot, Bhardwaj:2022maz, Bartsch:2022ytj, Mekareeya:2022spm}.\footnote{For $d=2$ these categorical defects have long been described in RCFTs \cite{Verlinde:1988sn,Petkova:2000ip}, a subject which has also seen a great deal of recent renewed interest \cite{Chang:2018iay,Komargodski:2020mxz,Thorngren:2019iar,Thorngren:2021yso}}

Higher dimensional generalizations of the well known Kramers-Wannier duality defects\footnote{See also \cite{Frohlich:2004ef,Frohlich:2006ch} for earlier works in 2d.} have recently attracted a great deal of attention starting from \cite{Choi:2021kmx,Kaidi:2021xfk}. It was first pointed out in \cite{Choi:2021kmx} that such defects appear when the theory $\cT$ is self-dual under some generalized gauging $\Phi$ of a discrete symmetry $G$:
\be
\cT / \Phi  \cong \cT  \, .
\ee
A symmetry defect $\fD$ can be constructed by performing the gauging operation on half spacetime and then using the isomorphism which implements the duality transformation to undo its action anywhere but on a very thin topological slab.
\bea    \begin{tikzpicture}        \filldraw[color=white, fill=white!85!blue] (0,-1) -- (0,1) -- (2,1) -- (2,-1) -- cycle;
		\draw[line width= 1.5, dashed] (0,-1) -- (0,1);
		\draw[line width=1.5] (2,-1) -- (2,1); 
		\node at (1,0) {$\cT/\Phi$};
		\node at (-0.5,0) {$\cT$};
		\node at (2.5,0) {$\cT$};
		\node at (3.25,0) {$=$};
		\draw[color=blue, line width = 1.5] (4,-1) -- (4,1);
		\node[below] at (2,-1) {$\Phi$};
		\node[below] at (0,-1) {$S$};
		\node[below] at (4,-1) {$\fD$};
    \end{tikzpicture}
\eea 
The symmetry thus obtained is not a standard zero-form symmetry, as the product with its inverse does not give back the identity, but rather a condensation defect \cite{Roumpedakis:2022aik} of the gauged symmetry.
\bea
&\fD \times \fD^\dagger = \cC^{G} \,  , \\
&\cC^G = \sum_{\gamma \in H^{p+1}(\Sigma, \, G) } \; U(\gamma) \; \exp\left(2 \pi i \varphi(\gamma)\right)\,,
\eea
where $U(\gamma)$ are the defects generating $G$ and $\varphi$ represent a choice of discrete torsion.
A striking example is $\cN=4$ SYM with gauge group $\SU(N)$, which is self dual under the gauging of its discrete $\bZ_N$ one-form symmetry at $\tau=i$. Similar reasoning also applies to different global forms of the SYM theory, with gauge groups $\PSU(N)_r$.
This can be further generalized to $\cN=4$ theories with different gauge groups\footnote{Due to the action of $S$-duality on the gauge algebra the methods described to construct duality defects work only in theories with simply laced algebras. The unique example in which such construction can be applied to the non-simply laced case is $B_2\cong C_2$ since the Langlands dual algebras are isomorphic.} \cite{Kaidi:2022uux} and to triality defects corresponding to the other fixed point $\tau = e^{2 \pi i /3}$ \cite{Choi:2022zal}.
A complete classification of the fusion algebra for these defects is however still missing (although the full fusion rules for the different global forms of $\mathfrak{su}(N)$ were computed in \cite{Antinucci:2022vyk} for $N$ prime). One of the aims of this work is to fill-in this gap.

Another very natural setting in which non-invertible defects may appear is $4d$ $\cN=2$ theories of class $\cS$ \cite{Gaiotto:2009we}, as noted in \cite{Bashmakov:2022jtl}. These theories are obtained as the dimensional reduction of a 6d $\cN=(2,0)$ SCFT on a Riemann surface $\Sigma_g$\footnote{In this work we only focus on theories of type $A_{N-1}$ in the absence of punctures. We expect many of our results to extend to the case of regular punctures, while we have nothing definite to say about the irregular ones. Furthermore for technical reasons we assume $N$ to be a prime number.
}, and have a conformal manifold (\ie \ a space of exactly marginal deformations) equal to the moduli space of complex structures of the Riemann surface, whose point we denote generically by $\Omega$. Moreover, these theories have a large one-form symmetry group, their global forms (\ie \ the set of their genuine line operators) are classified by Lagrangian lattices $\cL$ inside $H_1(\Sigma_g, \bZ_N)$ \cite{Tachikawa:2013hya,Bhardwaj:2021pfz,Bhardwaj:2021mzl} and also enjoy an extended duality group $\text{MCG}(\Sigma_g)$ given by the group of large diffeomorphisms of the underlying Riemann surface \cite{Gaiotto:2009we}.
The classification of non-invertible duality defects for these theories can be done in three steps:
\begin{enumerate}
    \item Find Riemann surfaces $\Sigma_g$ with a nontrivial automorphism group $G(\Omega)\subset \text{MCG}(\Sigma_g)$. These will be the self-dual loci for the class $\cS$ theory. This problem has been solved for high enough genus in the mathematical literature \cite{breuer2000characters}.
    \item Understand the action of the duality group $G(\Omega)$ on global variants, which are Lagrangian lattices in $H_1(\Sigma_g, \, \bZ_N)$.
    \item Study the action of discrete gauging operations $\Phi$ on half-space. These turn out to form a central extension of $\Sp{(2g, \, \bZ_N)}$, which we dub $\Sp{(2g, \, \bZ_N)}_T$. These operations have already been studied in detail in \cite{Choi:2022zal,Kaidi:2022uux} for the case of $\bZ_N$ one-form symmetry. We generalize and streamline their construction.
\end{enumerate}
We construct a duality defect $\fD_\cL^M$ by composing the duality action $M \in \Sp(2g,\bZ_N)$ with an appropriate topological manipulation $\Phi_\cL^M$ which restores the initial duality frame choice. We can then compute the full set of fusion rules and the action on the line operators of the theory. Interestingly, we find that from the action of $\fD_\cL^M$ on lines, we can define a property called \emph{rank}, which almost fixes the structure of the fusion algebra. Defects of rank zero are invertible, and we will use the notation $\cU_\cL^M$ to highlight this fact.

A second, complementary approach is to study a 5d Symmetry TFT for the topological operators in our theory. Using this method, the SCFT $\cT$ is expanded into a topological $(d+1)$-dimensional slab
\bea \label{fig: freedsandiwch}
	\begin{tikzpicture}[scale=1.75]
		\filldraw[fill=white!70!blue, opacity=0.5] (0,0) -- (1,0.25) -- (1,1.25) -- (0,1) -- cycle;
		\node at (0.5, 0.625) {$\cT$}; 
		\node at (1.5,0.625) {$=$};
		\begin{scope}[shift={(2,0)}]
				\draw (0,0) --(2,0); \draw (1,0.25) -- (3,0.25); \draw (1,1.25) -- (3,1.25); \draw (0,1) -- (2,1);
		\filldraw[fill=white!70!blue, opacity=0.5] (2,0) -- (3,0.25) -- (3,1.25) -- (2,1) -- cycle;	
		
			\filldraw[fill=white!70!red, opacity=0.5] (0,0) -- (1,0.25) -- (1,1.25) -- (0,1) -- cycle;	
		\node at (1.5,0.625) {\small $ \text{TQFT}_{d+1}$};
			\node at (0.5, 0.625) {$\cL$};
				\node at (2.5, 0.625) {$\cR$};  
		\end{scope}
	\end{tikzpicture}
\eea
with topological boundary conditions $\cL$ on the farther end specifying the global structure of $\cT$. The local dynamics is encoded in the other boundary $\cR$, which can be thought of as a relative version \cite{Freed:2012bs} of $\cT$. The bulk TQFT can be constructed explicitly when a holographic dual of $\cT$ is known. The symmetry TFT for duality defects in $\cN=4$ SYM has been recently derived in \cite{Antinucci:2022vyk,Kaidi:2022cpf}. The self-duality symmetry has a simple interpretation in terms of topological twist defects $D[M]$ for a bulk zero-form symmetry $G(\Omega)$. At special points in the gravitational moduli space, which correspond to self-dual SCFTs, the Symmetry TFT must be modified by gauging a subgroup of the zero-form symmetry. The twisted sectors then become liberated codimension-2 operators $\fD[M]$ which, when placed at the boundary, give rise to the codimension-1 duality defects of the SCFT:
\bea
\begin{tikzpicture}[scale=1.75]
 	\filldraw[fill=white!70!red, opacity=0.5] (0,0) -- (1,0.25) -- (1,1.25) -- (0,1) -- cycle;	
	 \draw (1,0.25) -- (3,0.25); \draw (1,1.25) -- (3,1.25); 
 \filldraw[color=green, fill=white!70!green, opacity=0.5] (1.25,0.125) --(1.25,1.125) -- (1.75,1.25) -- (1.75,0.25);
 \draw[color=green, line width=1.5] (1.25,0.125) -- (1.25,1.125);
		\filldraw[fill=white!70!blue, opacity=0.5] (2,0) -- (3,0.25) -- (3,1.25) -- (2,1) -- cycle;	
		\draw (0,0) --(2,0); \draw (0,1) -- (2,1);
\node at (1.5,0.625) {$M$};
  \node at (1.25,1.45) {$D[M]$};
  \node at (4,0.625) {\LARGE $\underset{G(\Omega)}{\overset{\text{gauging}}{\leadsto}}$};
  \begin{scope}[shift={(5,0)}]
      \filldraw[fill=white!70!red, opacity=0.5] (0,0) -- (1,0.25) -- (1,1.25) -- (0,1) -- cycle;	
	 \draw (1,0.25) -- (3,0.25); \draw (1,1.25) -- (3,1.25); 
  \draw[color=green, line width=2] (1.25,0.125) -- (1.25,1.125);
		\filldraw[fill=white!70!blue, opacity=0.5] (2,0) -- (3,0.25) -- (3,1.25) -- (2,1) -- cycle;	
    \node at (1.25,1.45) {$\fD[M]$};
    \draw (0,0) --(2,0); \draw (0,1) -- (2,1);
  \end{scope}
\end{tikzpicture}
\eea
For class $\cS$ theories, the gauged zero-form symmetry corresponds to a subgroup $G(\Omega)$ of the large diffeomorphisms of $\Sigma_g$ which is un-Higgsed at low energies.
Thanks to this description, it is possible to compute the fusion algebra for the non-invertible duality defects by carefully examining the composition laws for $D[M]$. We explicitly construct the Symmetry TFT and use this approach to confirm our previous results, thus also providing a highly non-trivial check for the holographic proposal of \cite{Antinucci:2022vyk}.

The structure of the fusion rules is rather simple to describe. The duality defects compose in a group-like manner (\ie  \ the fusion is graded by $G(\Omega)$) and the categorical structure shows up as either decoupled TQFT ``coefficients'' $\cN$ or condensation defects $\cC$:
\be
\fD^{M_1}_\cL \; \times \fD_\cL^{M_2} = \cN^{(1,2)} \; \cC^{\cA_{1,2}} \; \fD_\cL^{M_1 \; M_2}.
\ee
The TQFT coefficient can be chosen as decoupled minimal $\cA^{N, \cN^{(1,2)}}$ TQFTs \cite{Hsin:2018vcg} modulo congruence. For a given rank of $\cN^{(1,2)}$, there are only two classes for $N$ prime. The condensation instead refers to the higher gauging of a subgroup $\cA$ of the $(\bZ_N)^g$ one-form symmetry. 
The appearance of both of this structures follows from some rather simple observations regarding the rank of the defects participating in the fusion process. Interestingly, when the group is non-abelian, the categorical data is also non-commutative, in the sense that $\fD_\cL^{M_1} \times \fD_\cL^{M_2}$ and $\fD_\cL^{M_2} \times \fD_\cL^{M_1}$ can display different categorical structures. They are however consistent with associativity, albeit in a nontrivial way.

The paper is organized as follows: in Section \ref{sec: global variants}  we discuss in detail the classification of different global forms for the class $\cS$ theories we will study. In Section \ref{sec:Duality defects and discrete topological manipulations} we give a precise definition of the non-invertible duality defects and describe in detail the algebra of discrete topological manipulations $\Phi$, its action on global variants and the way in which it can be used to extract the fusion rules. In Section \ref{sec:rankQFT} we introduce the concept of \emph{rank} of a non-invertible duality defect, we describe how to compute it and how it can be used to (almost) fix the form of the fusion algebra.
In Section \ref{sec:The $5d$ Symmetry TFT}  we give an alternative method to extract such data from a 5d TQFT description, following the analysis of \cite{Antinucci:2022vyk}. In Section  \ref{sec: Examples} we give some explicit applications of our methods to low genus cases. We conclude in Section \ref{sec: conclusions} with open questions and prospects for future investigations. Various technical details and tables of some fusion rules can be found in the Appendices.\\[1em]

While this work was being completed \cite{Bashmakov:2022uek} appeared, in which non-invertible duality defects for class $\cS$ theories are also studied. Our work should be viewed as complementary to theirs as it mostly focuses on computing the fusion algebras of these defects. We thank the authors of \cite{Bashmakov:2022uek} for acknowledging our project in their manuscript prior to publication.

\section{Global variants and Lagrangian lattices} \label{sec: global variants}
\label{sec:Global variants and Lagrangian lattices}
We consider 4d gauge theories with semi-simple gauge algebra $\mathfrak{g}$. Let $\mathfrak{g}^*$ be the Langlands dual algebra, which is isomorphic to $\mathfrak{g}$ in the simply-laced cases. If we denote by $\widetilde{G}$ and $\widetilde{G}^*$ the simply connected groups with algebra $\mathfrak{g}$ and $\mathfrak{g}^*$ respectively, in absence of charged matter the full set of line operators is labelled by a lattice $\Gamma =Z(\widetilde{G})\times Z(\widetilde{G}^*)$, which comes with a natural non-degenerate antisymmetric pairing $\langle\, ,\, \rangle$. $Z(\widetilde{G})$ and $ Z(\widetilde{G}^*)$ are isomorphic and label respectively electric and magnetic charges. $\Gamma $ includes mutually non-local operators and therefore, as pointed out in  \cite{Gaiotto:2010be, Aharony:2013hda}, the set of genuine line operators $W_{l\in \cL}$ of the theory is specified by the choice of a maximal isotropic (i.e. Lagrangian) sublattice $\cL\subset \Gamma $.
The one-form symmetry instead is identified with
\begin{equation}
    \cS=\Gamma /\cL \ .
\end{equation}
$\cS$ can also be understood as the set labelling the non-genuine line operators $T_{s\in \cS}$, which live in the twisted sectors of the one-form symmetry. The reason why $\cS$ is a quotient is that by adding a genuine line to a non-genuine one, the resulting line  remains in the same twisted sector. We will focus on the case in which $Z(\widetilde{G})$ does not have non-trivial proper subgroups. Then the exact sequence $1\xrightarrow[]{}\cL\xrightarrow{}\Gamma \xrightarrow{} \cS\xrightarrow[]{}1$ splits, so it is always possible to choose a  representative of $\cS$ (by abuse of notation we call it $\cS$ itself) which is also Lagrangian, and such that
\begin{equation}
    \Gamma =\cL \oplus \cS  \ .
\end{equation}
Since we will study class $\cS$ theories, we will be mostly interested in theories with charged matter which can partially screen the line operators. In this case $Z(\widetilde{G})\cong Z(\widetilde{G}^*)$ is replaced by a quotient $Z(\widetilde{G})/\Lambda$, where $\Lambda \subset Z(\widetilde{G})$ is the subgroup of charges screened by matter. Consider a class $\cS$ theory of type $\mathfrak{g}$, obtained by compactifying the 6d $\cN=(2,0)$ theory of type $\mathfrak{g}$ on a genus $g$ Riemann surface $\Sigma _g$ without punctures. The corners of the conformal manifold with a quiver description correspond to a pair of pants decomposition in terms of $2g-2$ three-punctured spheres glued by $3g-3$ very long tubes. The Lagrangian is written in terms of $3g-3$ $\cN=2$ vector multiples of $\mathfrak{g}$ coupled to $2g-2$ copies of the $T_{\mathfrak{g}}$ theory, namely tri-fundamental hypermultiplets, corresponding to the three-punctured spheres \cite{Gaiotto:2009we}. Denote by $\widetilde{G}_0$ the simply connected group with algebra $\mathfrak{g}$, so that $\widetilde{G}=\widetilde{G}_0^{3g-3}$. Each tri-fundamental is charged with respect to the diagonal $Z(\widetilde{G}_0)$ of the three vector multiplets coupled to it. Since each vector multiplet is coupled to exactly two $T_{\mathfrak{g}}$ and $\Sigma _g$ has no puncture, the diagonal of all the $Z(\widetilde{G}_0)$ charges of the hypermultiplets is not acted upon by the center symmetry. This means that $\Lambda$ has co-dimension one in $Z(\widetilde{G}_0)^{2g-2}$, and the set of unscreened electric charges is $Z(\widetilde{G}_0)^g$. 

The bottom line is that the classifying lattice for global variants of the class $\cS$ theory is $\Gamma =Z(\widetilde{G}_0)^{2g}$, which coincides with $H^1(\Sigma _g, Z(\widetilde{G}_0))$ \cite{Tachikawa:2013hya, Bhardwaj:2021pfz}. 

In the case of $\mathfrak{g}=A_{N-1}$, we have $Z(\widetilde{G}_0)=\bZ _N $ and $\Gamma =(\bZ _N)^{2g}$. The pairing on $\Gamma$ is given by  
\begin{equation}
    \langle v, u\rangle =v^{\sT}\cJ u \ \ , \ \ \ \ \cJ=\left(\begin{array}{cc}
      0   & \unit _g \\
     -\unit  _g  & 0 
    \end{array}\right) \ .
\end{equation}
In this paper we will restrict to the case where $N$ is a prime number. Then a Lagrangian lattice $\cL$ specifying a global variant corresponds to the choice of $g$ linearly independent vectors $v_1,...,v_g\in (\bZ _N)^{2g}$ such that 
\begin{equation}
    v_i^{\sT}\cJ v_j=0 \ \ , \ \ \ \forall i,j =1,...,g  \ .
\end{equation}
We will label these lattices by $2g\times g$ matrices $\cL =(v_1,...,v_g)$ of rank $g$, which satisfy $\cL ^{\sT}\cJ \cL=0$. One such matrix is 
\begin{equation}
    \cE=\left(\begin{array}{c}
        \unit _g  \\
          0
    \end{array}\right)
\end{equation}
and we will call the corresponding theory the \emph{electric variant}. All the others are obtained by acting on $\cE$ with matrices $M\in \Sp(2g,\bZ _N)$. The action on $\Sp(2g,\bZ _N)$ on Lagrangian lattices $\cL$ is transitive, and the stabilizer is isomorphic to the group of symplectic matrices leaving  $\cE$ invariant up to a change of basis. These matrices are of the form
\begin{equation}
    \left(\begin{array}{cc}
       u  & us  \\
       0  & {u^{\sT}}\inv
    \end{array}\right) \  , \ \ \  \ \ \ u\in \GL(g,\bZ _N) \ , \ \ s^{\sT}=s \ ,
\end{equation}
and generate the parabolic subgroup $\cP (2g,\bZ _N) \subset \Sp(2g,\bZ _N)$. 
We conclude that the global variants are labelled by the right coset $\Sp(2g,\bZ _N)/\cP (2g,\bZ _N)$, and therefore their number is\footnote{We use that $|\Sp(2g,\bZ _N)|=N^{g^2}\prod _{k=1}^g (N^{2k}-1)$ and $|\cP(2g,\bZ _N)|=N^{\frac{g(g+1)}{2}}\prod _{k=0}^{g-1} (N^g-N^k)$.}
\begin{equation}
\label{eq:number global variants}
    N_{\text{global variants}}
    =\prod _{k=0}^{g-1}(N^{k+1}+1) \ .
\end{equation}
Note that for $g=1$ we obtain $N+1$, which is indeed the number of global variants of $\mathfrak{su}(N)$ YM theories for $N$ prime, including the electric variant $\SU(N)$ and the $N$ magnetic variants $\PSU(N)_r$, $r=0,...,N-1$. In this case all other variants can be reached from the electric $\SU(N)$ variant by gauging the $\bZ_N$ one-form symmetry with an appropriate discrete torsion  \cite{Kapustin:2014gua, Gaiotto:2014kfa}. In order to extend this idea to generic $g$ we rewrite \eqref{eq:number global variants} using the q-binomial theorem as
\begin{equation}
\label{eq:q-binomial}
    N_{\text{global variants}}=\sum _{k=0} ^g \left(\begin{array}{c}
          g \\
          k
    \end{array}\right)_N N^{\frac{k(k+1)}{2}} \ .
\end{equation}
Here we introduced the Gaussian binomial coefficient
\begin{equation}
    \left(\begin{array}{c}
          g \\
          k
    \end{array}\right)_N=\frac{(1-N^g)(1-N^{g-1})\cdots (1-N^{g-k+1})}{(1-N)(1-N^2)\cdots (1-N^k)} \, 
\end{equation}
which also counts the number of $(\bZ _N)^k$ subgroups of $(\bZ _N)^g$. 
After this manipulation, equation \eqref{eq:q-binomial} has a clear interpretation as the number of inequivalent ways to gauge a $(\bZ_N)^k$ subgroup of the $(\bZ_N)^g$ one-form symmetry with possible discrete torsion, which for $N$ prime is encoded in a $k \times k$ symmetric matrix.

Another convenient way to label the global variants, also used in \cite{Bashmakov:2022uek}, is to use $2g \times 2g$ symplectic matrices $V$ instead, subject to the identification $V\sim VP$, $P\in \cP(2g,\bZ _N)$. 
Writing the symplectic matrices in block form
\begin{equation}
    V=\left(\begin{array}{cc}
        A & B \\
        C & D
    \end{array}\right)\ \ , \ \ \ A^{\sT}C-C^{\sT}A=B^{\sT}D-D^{\sT}B=0 \ , \ \ \ A^{\sT}D-C^{\sT}B=\unit_g 
\end{equation}
the Lagrangian lattice labelling the global variant is 
\begin{equation}
    \cL =\left(\begin{array}{c}
         A \\
          C
    \end{array}\right) 
\end{equation}
with the identification $A\sim Au$, $C\sim Cu$ where $u\in \GL(g,\bZ _N)$. This corresponds to the right action on the $2g\times g$ matrix $\cL$ as $\cL\rightarrow \cL u$. The condition $A^{\sT}C=C^{\sT}A$ is precisely the requirement $\cL ^{\sT}\cJ\cL=0$ that the genuine lines are mutually local. This way of labelling the global variants also makes explicit the choice of representative for $\cS=\Gamma /\cL$. Indeed this is just
\begin{equation}
\cS=\left(\begin{array}{c}
   B   \\
   D   
\end{array}\right) \,. 
\end{equation}
The equation $B^{\sT}D=D^{\sT}B$ implies that $\cS$ is also Lagrangian and the condition $A^{\sT}D-C^{\sT}B=\unit_g $ is nothing but  $\cL ^{\sT}\cJ \cS=\unit _g$, namely the fact that the two have canonical pairing. Notice that in terms of $\cL$ and $\cS$ we have $V=(\cL\, | \, \cS)$.

We will see shortly that there is a natural mapping between this parametrization and the inequivalent ways of gauging the one-form symmetry with a choice of discrete torsion. Roughly speaking $C$ will encode the information about the choice of gauged subgroup, while the choice of discrete torsion is encoded in $A$.

\section{Duality defects and discrete topological manipulations}
\label{sec:Duality defects and discrete topological manipulations}
In this Section, we define generic duality defects in theories with a $(\bZ_N)^g$ one-form symmetry and describe their composition properties.
In theories of class $\cS$, the duality group has a natural $\Sp{(2g, \bZ_N)}$ action\footnote{Geometrically, this follows from the short exact sequence 
\be
\tor \rightarrow \MCG(\Sigma_g) \rightarrow \Sp{(2g, \bZ)}
\ee
where $\MCG(\Sigma_g)$ is mapping class group and $\tor$ is the Torelli group.}
on $H_1(\Sigma_g, \bZ_N)$ which sends a Lagrangian lattice $\cL$ to
\be
\cL \to M \cL \, ,  \ \ \ \ \ \ \ M=\left(\begin{array}{cc}
A     & B \\
C     & D
\end{array}\right) \in \Sp{(2g, \bZ_N)} \, 
\ee
while acting on the complex structure matrix as $\Omega \to M(\Omega) = (A \Omega + B)(C \Omega + D)\inv$. 

As remarked in \cite{Choi:2021kmx, Kaidi:2022uux} in the case of $g=1$, the same action on global variants can also be realized by appropriately choosing a topological manipulation $\Phi_\cL^{M}$. This corresponds to gauging a subgroup $\cA$ of the one-form symmetry, possibly with discrete torsion. The space of topological manipulations forms a central extension of $\Sp{(2g, \bZ_N)}$, which we denote $\Sp{(2g, \bZ_N)}_T$. When we represent a topological manipulation $\Phi_\cL^{M}$ as a matrix in $\text{Sp}(2g, \mathbb{Z}_N)$, it acts on the global variant parameterized by the matrix $V$ on the right. We observe that, similar to the duality transformations acting from the left, there exists a parabolic subgroup $\mathcal{P}(2g, \mathbb{Z}_N)$ of $\text{Sp}(2g, \mathbb{Z}_N)$ for topological manipulations, which leaves the global variant unchanged. Importantly, the right action of these topological manipulations ensures that they commute with the duality action.

Given a point $\Omega$ on the conformal manifold $\cM$ stabilized by a subgroup $G(\Omega) \subset \text{MCG}(\Sigma_g)$ and a choice of global variant $\cL$, we can define the duality defect $\fD_\cL^M$, $M \in G(\Omega)$ by composing a duality transformation with a topological manipulation\footnote{Our conventions are that defects act on operators on their right.}
\bea    \begin{tikzpicture}    \node[left] at (-1,0) {$\ds  \fD_\cL^M = M \circ \Phi_\cL^{M\inv} $};
    \filldraw[color=white, fill=white!80!blue] (2,-1) -- (2,1) -- (4,1) -- (4,-1) -- cycle;
    \draw[color=black, line width = 1.5, dashed] (2,-1) -- (2,1);
     \draw[color=black, line width = 1.5] (4,-1) -- (4,1);
     \node at (1,0) {$\cL$};
     \node at (3,0) {$M\inv\cL$};
     \node at (5,0) {$\cL$};
     \node[below] at (4,-1) {$\ds \Phi_\cL^{M\inv}$};
     \node[below] at (2,-1) {$M$};
     \node at (6,0) {$=$};
     \draw[color=blue, line width=3] (7,-1) -- (7,1);
     \node[below] at (7,-1.2) {$\ds \fD_\cL^M$};
    \end{tikzpicture}
\eea
The fusion $\fD_\cL^{M_2} \times \fD_\cL^{M_1}$ between duality defects can be understood from the compositions laws for the topological manipulations $\Phi_\cL^M$ on half space. After expanding both defects into slabs we slide the duality transformation $M_1$ across $\Phi_\cL^{M_2\inv}$ as shown below:\footnote{To avoid clutter we leave implicit the labelling $\cL$ of the original theory. We also use the shorthand $M_{2,1}= M_2 M_1$.}
\bea
\begin{tikzpicture}     \filldraw[color=white, fill=white!80!blue] (0,-1) -- (0,1) -- (2,1) -- (2,-1) -- cycle;
    \draw[color=black, line width = 1.5, dashed] (0,-1) -- (0,1);
     \draw[color=black, line width = 1.5] (2,-1) -- (2,1);
      \node at (1,0) {$M_2\inv\cL$};
       \node[below] at (2,-1) {$\ds \Phi_\cL^{M_2\inv}$};
     \node[below] at (0,-1) {$M_2$};
     \begin{scope}[shift={(3,0)}]
     \filldraw[color=white, fill=white!70!pink] (0,-1) -- (0,1) -- (2,1) -- (2,-1) -- cycle;
    \draw[color=black, line width = 1.5, dashed] (0,-1) -- (0,1);
     \draw[color=black, line width = 1.5] (2,-1) -- (2,1);
      \node at (1,0) {$M_1\inv\cL$};
       \node[below] at (2,-1) {$\ds \Phi_\cL^{M_1\inv}$};
     \node[below] at (0,-1) {$M_1$};
     \node at (2.5,0) {$=$};
     \end{scope}
     \begin{scope}[shift={(6,0)}]
      \filldraw[color=white, fill=white!80!blue] (0,-1) -- (0,1) -- (1.5,1) -- (1.5,-1) -- cycle;
      \filldraw[color=white, fill=white!80!purple] (1.5,-1) -- (1.5,1) -- (3,1) -- (3,-1) -- cycle;
      \filldraw[color=white, fill=white!70!pink] (3,-1) -- (3,1) -- (4.5,1) -- (4.5,-1) -- cycle;
      \draw[color=black, line width = 1.5, dashed] (0,-1) -- (0,1);
      \draw[color=black, line width = 1.5, dashed] (1.5,-1) -- (1.5,1);
       \draw[color=black, line width = 1.5] (3,-1) -- (3,1);
        \draw[color=black, line width = 1.5] (4.5,-1) -- (4.5,1);
        \node at (0.75,0) {$M_2\inv \cL$};
          \node at (2.25,0) {$M_{2,1}\inv \cL$};
            \node at (3.75,0) {$M_1\inv \cL$};
        \node[below] at (4.5,-1) {$\ds \Phi_\cL^{M_1\inv}$};
          \node[below] at (3,-1) {$\ds \Phi_{M_1\inv \cL}^{M_{2,1}\inv  M_1}$};
          \node[below] at (1.75,-1) {$M_1$};
          \node[below] at (0,-1) {$M_2$};
          \node at (5.25,0) {$=$};
     \end{scope}
     \begin{scope}[shift={(12,0)}]
       \filldraw[color=white, fill=white!80!purple] (0,-1) -- (0,1) -- (3,1) -- (3,-1) -- cycle;
        \draw[color=black, line width = 1.5,dashed] (0,-1) -- (0,1);
              \draw[color=black, line width = 1.5] (3,-1) -- (3,1);
              \node at (1.5,0) {$M_{2,1}\inv \cL$};
       \node[below] at (0,-1) {$M_{2,1}$};
       \node[below] at (2.25,-1) {$\ds \Phi_{M_1\inv \cL}^{M_{2,1}\inv M_1} \circ \Phi_\cL^{M_1\inv}$};
     \end{scope}
    \end{tikzpicture}
\eea
Since the left duality action and the right topological actions commute we have\be
\Phi_{M_1\inv \cL}^{M_{2,1}\inv M_1} = \Phi_\cL^{M_2\inv} \, .
\ee 
We will now discuss the structure of $\Sp{(2g,\bZ_N)}_T$, starting with the example of $g=1$ to then move onto the more general case.

\subsection{Topological manipulations for $\bZ_N$: $\SL{(2,\bZ_N)}_T$}
Let us review the structure of $\SL{(2, \bZ_N )}_T$ \cite{Gaiotto:2014kfa,Bhardwaj:2020ymp,Choi:2022zal}, the space of topological manipulations for a $\bZ_N$ symmetry. 
The generators of these manipulations are:
\begin{equation}
\begin{aligned}
     \sigma &: \ &&\left[ \sigma Z\right](B) = \frac{1}{|H^2(X,\bZ_N)|^{1/2}} \sum_{b \in H^2(X,\bZ_N)} \exp\left( \frac{2 \pi i}{N} \int b \cup B \right) \; Z(b) \, , \\
\tau(k)&: \ &&\left[ \tau(k) Z\right](B) = \exp\left( \frac{2 \pi i k}{2 N} \int \fP(B) \right) \; Z(B) \, , \\
\nu(u) &: \ && \left[ \nu(u) Z\right](B) = Z(u B) \, \ \ \ \ u \in \bZ_N^\times , 
\end{aligned}
\label{eq: SL2 manip}
\end{equation}
$\fP$ being the Pontryagin square operation $\fP : H^2(X, \, \bZ_N) \to H^4(X, \, \bZ_{\gcd(2,N)N})$ (see \eg{} \cite{Benini:2018reh}). Strictly speaking the discrete gauging $\sigma$ maps the $\bZ_N$ 1-form symmetry to its Pontryagin dual $(\bZ_N)^* \cong \bZ_N$. Since we want to perform successive gauging procedures, we will always implicitly use this isomorphism. Throughout the rest of the paper we will often omit the overall normalization factor for the discrete gauging.
These operations can be also represented as matrices in $ \SL{(2,\bZ_N)}_T$
\be \label{eq: SL2 matrices}
\sigma = \mat{0 & -1 \\ 1 & 0} \, , \ \ \
\tau(k) = \mat{1 & k \\ 0 & 1} \, , \ k \in \bZ_N \ \ \
\nu(u) = \mat{u\inv & 0 \\ 0 & u} \, , \ u \in \bZ_N^\times \, 
\ee
acting on the matrix $V = \left( \cL \; \vert \; \cS \right)$ on the right.
The transformations $\tau(k)$ and $\nu(u)$ do not alter the global structure, \ie \  they leave $\cL$ invariant. $\nu(u)$ corresponds to a different choice of basis in the space of lines, while $\tau(k)$ amounts to a background discrete theta angle. 
Together they form the parabolic subgroup $\cP(2,\bZ_N)$ of $\SL{(2,\bZ_N)}$.
The algebra of these transformations can be computed straightforwardly. The most interesting relation, which we call the \emph{``K-formula''} (see appendix \ref{app: composition comp} for a derivation), follows from considering a two-fold gauging process and reads
\be
\sigma \; \tau(k) \; \sigma = Y_k \; \nu(-k\inv) \; \tau(-k) \; \sigma \; \tau(-k\inv) \, .
\ee
where 
\be
Y_k = \sum_{b \in H^2(X, \, \bZ_N)} \exp\left( \frac{2 \pi i k}{2 N} \fP(b)  \right)
\ee
is an invertible $4d$ two-form gauge theory \cite{Hsin:2018vcg}. If instead we have no intermediate torsion
\be
\sigma \; \sigma = \nu(-1) \, .
\ee
Furthermore, it holds that $\sigma  \nu(u) = \nu(u\inv) \sigma$.
Thus, whenever we have subsequent $\sigma$ insertions, we can use the K-formula to reduce their number. Repeating this process we can bring every element of $\Phi \in \SL(2,\bZ_N)_T$ into the \emph{Standard form}:
\be
\Phi = P(u,s) \; \sigma \; \tau(k) \, , \ \ \ P(u,s) = \nu(u) \; \tau(s) \, .
\ee
Henceforth topological manipulations are always assumed to be in the standard form.

\paragraph{Action on global variants.} Let us briefly discuss the action of $\SL{(2,\bZ_N)}_T$ on global variants. This will be used to set up a precise dictionary between the matrix $\cL$ and the discrete operations $\Phi$. Let
\be
\Phi_k = \sigma \; \tau(k) \, 
\ee
and $\cL$ be the chosen variant for our 4d gauge theory. This has both genuine lines $W_l \, , l \in \cL$ and twisted sector lines $T_s \, , s \in \cS$ which are attached to open one-form symmetry surfaces $U_s$. 
Our aim is to understand the spectrum of genuine lines after applying $\Phi_k$. This amounts to classify which lines of the ungauged theory are invariant under background $\bZ_N$ gauge transformations
\be
B \to B + \delta \lambda \, ,
\ee
in the presence of a background discrete theta angle $e^{\frac{2 \pi i k}{2 N} \fP(B)}$. 
Genuine lines $W_l$ are charged under the one-form symmetry and they will pick up a phase $e^{\frac{2 \pi i}{N} l \int \lambda } = e^{- \frac{2 \pi i}{N} l \int \lambda \cup \text{PD}(\gamma)}$. Due to the presence of the discrete theta angle non-genuine lines $T_s$ are not invariant either. Their insertion on a curve $\gamma$ corresponds to a background $B$ which fulfills $ \delta B = s \text{PD}(\gamma)$.
The discrete torsion term then fails to be gauge invariant by a phase $e^{\frac{{2 \pi i k}}{N} s \int \lambda \cup \text{PD}(\gamma) }$.
Thus gauge invariant operators are generated by the dyonic line $D_{k, \, 1}\equiv W_k \; T_1$: the generator of the new Lagrangian lattice $\Phi_k \cdot \cL$ is $k \cL + \cS$.

\paragraph{Fusion rules.}
To derive the fusion rules, we analyze two subtleties involving the half-space composition in $\SL{(2, \; \bZ_N)}_T$.
The first comes from the invertible theories $Y_k$. While on a closed manifold, their partition function is just a phase; on a manifold $X^+$ with boundary $\partial X^+$, the TQFT $Y_k$ becomes the anomaly-inflow theory for a 3d TQFT\cite{Hsin:2018vcg}. This theory is not completely determined by the anomaly alone; however, a minimal choice always exists and is given by the minimal TQFT $\cA^{N, k}$.\footnote{The sign of $k$ depends on conventions for the orientation.} The composite system $Y_k(X^+) \times \cA^{N, k}(\partial X^+)$ is anomaly-free and well-defined. Thus, on half-space, we should interpret the appearance of $Y_k$ as an indicator of the presence of a decoupled 3d TQFT $\cA^{N, k}$. This gives our first rule:
\be  \label{eq: rule1}
\boxed{\begin{array}{ccc}
 & Y_k \ \text{on half space} \ X^+ = \text{decoupled TQFT} \ \cA^{N,k} \ \text{on} \ \partial X^+  & 
\end{array}}
\ee
The second subtlety has to do with the appearance of condensates $\cC^{\bZ_N}$ for the one-form symmetry \cite{Roumpedakis:2022aik}. As explained in \cite{Choi:2022zal} condensates appear when two gauging operations compensate each other in half-space. For instance consider $\sigma \, \sigma \, \tau(k)$ on $X^+$
\be
\left[\sigma \, \sigma \, \tau(k) Z\right] (B) = \sum_{b, \, c \in H^2(X^+, \bZ_N)} \exp\left( \frac{2 \pi i}{N} \int b  \cup (c + B) + \frac{k}{2} \fP(c)  \right) \; Z(c) \,. 
\ee
Naively, we could integrate out $b$ to enforce $c = -B$. However, on half-space, this is no longer true. As in the previous case, the theory for the $b$ field is inconsistent on its own in the presence of a boundary. To recover gauge invariance, the minimal choice is to add a 3d $\bZ_N$ gauge theory with DW twist $\alpha = - k N$. Its gauge field $a$ will couple to the bulk one-form symmetry. Now the integral on half-space can be performed, leaving behind a boundary term:
\be
\cC^{\bZ_N} = \sum_{\gamma \in H_1(\partial X^+, \bZ_N)} U(\gamma) \; \exp\left( \frac{2 \pi i k}{2} \int \text{PD}(\gamma) \; \beta \left(\text{PD}(\gamma) \right)\right) \, ,  \ \ \ \text{PD}(\gamma) = a
\ee
with $\beta$ the Bockstein map: $H^1(Y, \bZ_N) \to H^2(Y, \bZ_N)$.\footnote{Since we work with $N$ odd and on spin manifold, such DW term is always trivial and we can safely forget about this label in condensates.} 
This is exactly the condensation defect. Thus the second rule is:
\be \label{eq: rule2}
\boxed{\begin{array}{ccc}
 & \text{Un-gauging} \ \bZ_N \ \text{on half space} \ X^+ = \text{condensation defect} \ \cC^{\bZ_N} \ \text{on} \ \partial X^+  & 
\end{array}}
\ee

\paragraph{Example: Triality defects.}  We give a sample computation for the fusion of defects $D_\cL^{M_1} \times D_\cL^{M_2}$. For illustrative purposes we consider the triality defect $\fD_\cL^{ST}$ which appears in $\cN=4$ $\mathfrak{su}(N)$ SYM at $\tau = e^{\frac{2 \pi i}{3}}$.
We start by determining the topological manipulations $\Phi_\cL^{M\inv}$ mapping $V$ to $M\inv V$. This can be done by parametrizing $V =  \cE  \; \tau(k_\cL) \; \sigma$, for some known $k_\cL$. Clearly $\Phi_{\cL}^{M\inv}= V\inv M\inv V$. Setting $M^{-1} V = \tau(k_{M\inv \cL}) \; \sigma \; P $ for some parabolic element $P$ the K-formula gives
\be
\Phi_\cL^{M\inv}  \overset{\text{Standard}}{=} \nu(q\inv) \; \tau(-q) \; \sigma \; \tau(- q\inv)  \; P  \, , \ \ q = k_{M\inv \cL} - k_\cL \, .  
\ee
If instead $q=0$ then $\Phi_\cL^{M\inv}$ is a parabolic element and the global structure is left invariant $M\inv \cL = \cL$, giving rise to an invertible defect. Having computed both $\Phi_\cL^{M_1\inv}$ and $\Phi_{M_1\inv \cL}^{M_{2,1}\inv \; M_1}$ the fusion rules are obtained by applying the rules \eqref{eq: rule1}, \eqref{eq: rule2}.
For $M_1 = M_2 = ST$ the transformations are
\bea
&\Phi_{\cL}^{(ST)\inv} = \ \begin{cases} \sigma \, , \ &\cL = (1,0)^\sT \\  \;\tau(-a) \; \sigma \; \tau(-a\inv) \; \nu(-a) \, , \ a=-r\inv(1 + r + r^2) \, , \  &\cL = (r,1)^\sT \end{cases}  \ \ \ \\ 
&\Phi_{(ST)\inv\cL}^{(ST)\inv} = \begin{cases}  \tau(-1) \; \sigma \; \tau(-1) \; \nu(-1)  \, , \ &\cL = (1,0)^\sT \\  \; \tau(-b) \; \sigma \; \tau(-b\inv) \; \nu(-b)\, , \ b=(r+1)\inv r\inv (r^2 + r + 1) \, , \ &\cL = (r,1)^\sT \end{cases}
\eea
acting on $V$ on the right.
These lead to the fusion rules\footnote{The TQFT coefficient comes from $-a - a^2 b\inv = (r^2 + r + 1) r\inv (1 - r - 1) = -(r^2 + r +1)$. This matches previous results using 5d techniques \cite{Antinucci:2022vyk}. }

\bea
\fD_\cL^{ST} \times \fD_{\cL}^{ST} &=  \cA^{N,-(r^2 + r +1)} \times \fD_\cL^{(ST)^2} \, , \ \ &&\text{if} \ r^2 + r + 1 \neq 0 \mod N \\ \cU_{\cL}^{ST} \times \cU_{\cL}^{ST} &=  \cU_\cL^{(ST)^2} \, , \ \ &&\text{if} \ r^2 +r + 1=0 \mod N   
\eea

\subsection{Topological manipulations for $(\bZ_N)^g$ : $\Sp{(2g, \; \bZ_N)}_T$}
We now generalize our analysis to the group $\Sp{(2g, \bZ_N)}_T$ describing topological manipulations for 4d theories with $(\bZ_N)^g$ one-form symmetry. These are the relevant ones for the description of self-duality defects in $A_{N-1}$ theories of class $\cS$. As many features mirror the previous discussion we will be brief and relegate long computations to Appendix \ref{app: composition comp}.
In what follows we will be gauging subgroups $\cA$ of the one-form symmetry $(\bZ_N)^g$. Let us start by setting up some notation.
We describe $\cA$ as a sub-lattice of $(\bZ_N)^g$. Since $N$ is a prime number $\cA$ is isomorphic to $(\bZ_N)^r$ for some $r$. We will call $r$ the rank of the lattice. $\cA$ can be specified by choosing $r$ generators $\cA_1$ ... $\cA_r$ which we package into a rectangular matrix
\be
C_\cA = \mat{ \cA_1 & ... & \cA_r } \, .
\ee
This description is redundant since $C_\cA u$, $u \in \GL{(r , \bZ_N)}$ describes the same lattice. 
The number of distinct lattices of rank $r$ is given by the q-binomial coefficient $\binom{g}{r}_N$. Given a lattice $\cA$ we denote by $\Tilde{\cA}$ the quotient $(\bZ_N)^g/\cA$. Since the sequence $1\rightarrow \cA \rightarrow (\bZ_N)^g \rightarrow \Tilde{\cA}\rightarrow 1$ splits, we get $(\bZ_N)^g = \cA \times \Tilde{\cA}$. We will make a choice for the splitting by specifying a second matrix $C_{\Tilde{\cA}}$.
We define duals $C_\cA^*$ and $C_{\Tilde{\cA}}^*$ by the equations
\be
C_\cA^* \; C_\cA = \unit_{r} \, , \ \ C_{\Tilde{\cA}}^* \; C_{\Tilde{\cA}} = \unit_{g-r} \, , \ \ C_\cA^* \; C_{\Tilde{\cA}} = C_{\Tilde{\cA}}^* \; C_\cA = 0 \, .
\ee
The completeness relation reads $\unit_g = C_\cA \; C_\cA^* + C_{\Tilde{\cA}}\; C_{\Tilde{\cA}}^*$. We also denote the ``join'' and ``meet'' operations on lattices by $\cA \smallvee \cB$ and $\cA \smallwedge \cB$ respectively \footnote{The join of two lattices is defined as the span of union while the meet is the intersection.}.

Given a linear map $M: (\bZ_N)^g \to (\bZ_N)^g$ we define its restriction onto $\cA$ by
\be
M_\cA = C_\cA^\sT \; M C_\cA \, 
\ee
and a lift back to the original space by
\be
{M_\cA}^\cA = (C_\cA^*)^\sT \;  M_\cA \; C_\cA^* \, ,
\ee
such that $((M_\cA)^\cA)_\cA = M_\cA$. Given two sub-lattices $\cA$ and $\cB$ with trivial meet we also define double restrictions
\be
M_{\cA \cB} = C_\cA^\sT M C_\cB \equiv (M_{\cB \cA})^\sT \, ,
\ee
and double lifts
\be
(M_{\cA \cB})^{\cA \cB} = (C_\cA^*)^\sT \; M_{\cA \cB} \; C_{\cB}^* \, .
\ee
An (overcomplete) set of generators for $\Sp{(2g, \, \bZ_N)}_T$ is given by
\bea \label{eq: discrete manip Zg}
&\sigma(C_\cA) : \ &&\left[ \sigma(C_\cA) Z \right](B) =  \frac{1}{\vert H^2(X, \cA) \vert^{\frac{1}{2}}}\sum_{b_\cA \in H^2(X, \cA )} e^{\frac{2 \pi i}{N} \int b_\cA \cup B_\cA} \; Z( C_\cA b_\cA + C_{\Tilde{\cA}} B_{\Tilde{\cA}}  ) \, ,  \\
&\tau(S) : \ && \left[ \tau(S) Z \right](B) = \exp\left( \frac{2 \pi i}{2 N}\int \fP^S(B) \right) \; Z(B) \, , \ \ \\
&\nu(U) : \ &&  \left[ \nu(U) Z \right](B) = Z(U B) \, ,  \ \ U \in \GL{(g, \bZ_N)} \, 
\eea
where the (generalized) Pontryagin square $\fP^S$ is defined as
\be
\fP^S(B) = \sum_i S_{ii} \; \fP(B_i) + 2 \sum_{i>j} S_{ij} \; B_i \cup B_j \, .
\ee
The minimal coupling $b_\cA \cup B_\cA$ follows from considering the cup product on $(\bZ_N)^g$: $b \cup B \equiv B^* \cup b$. We restrict $b= C_\cA b_\cA$ and expand $B^* = B_\cA^* C_\cA^* +  B_{\Tilde{\cA}}^* C_{\Tilde{\cA}}^*$. We then declare $B_\cA^*$ to be the new $\cA$ background $B_\cA$. With these conventions, gauging $\cA$ twice leads back to the original theory up to charge conjugation on $C_\cA$.\footnote{On the other hand the standard isomorphism $(V^*)^*=V$ gives a gauging operation such that $\sigma(C_\cA) \sigma({C_\cA^*}^\sT)$ leads back to the original theory. The two differ by left composition with a $\nu(U)$ transformation.}
The transformations \eqref{eq: discrete manip Zg} correspond to matrices $M$ in $\Sp{(2g, \bZ_N)}$
\be
\sigma(E_j) = \mat{ \unit - E_j & - E_j \\ E_j & \unit - E_j} \, \ \ \ \tau(S) = \mat{\unit & S \\ 0 & \unit} \, \ \ \ \nu(U) = \mat{{U\inv}^\sT & 0 \\ 0 & U }
\ee
with 
\begin{equation}
    (e_j)_k = \delta_{jk} \ \ \ \text{and}   \ \ \  E_j = \left( 0 \; ...  \; \right.  \underbrace{ e_j }_{j} \left.  \; ... \; 0\right) \ .
\end{equation}
We discuss the representation theory of generic transformations in Appendix \ref{app: rep}. As before the matrices $\tau(S)$ and $\nu(U)$, which do not change the global structure of the theory, generate the parabolic subgroup $\cP(2g,\bZ_N)$ of $\Sp{(2g, \bZ_N)}$.

We would now like to generalize the \emph{``K-formula''} in this setting. We consider a double gauging
\be
\sigma(C_\cA) \;\tau(S) \; \sigma(C_\cB) \, .
\ee

For simplicity we assume that $S = {S_{\cA}}^\cA$ restricts to $\cA$. The derivation as well as the general case is given in Appendix \ref{app: composition comp}. We introduce the meet $\cC = \cA \smallwedge \cB $, the disjoint lattices $\cA' = \cA/\cC$, $\cB' = \cB / \cC$ and the corresponding generators and dual generators, with obvious notation. We find:
\be\label{eq: K-formula g}
\sigma(C_\cA) \; \tau({S_\cA}^\cA) \; \sigma(C_\cB) = Y_{S_{\cC}} \; \tau(- {(S\inv)_\cC}^\cC \,) \; \nu(U_{\cA' \, \cB' \, \cC}) \; \sigma\left( C_{\cA'} \vert C_{\cB'} \vert C_\cC \right) \; \tau(-X_{\cA' \, \cB' \, \cC})
\ee
where
\bea
&U_{\cA' \, \cB' \, \cC} = C_{\cA'} \left( C_{\cA'}^* - S_{\cA' \; \cC} \; S_\cC\inv C_{\cC}^*  \right) - C_\cC \; S_\cC\inv \; C_\cC^* + C_{\cB'} \; C_{\cB'}^* + \Tilde{C} \; \Tilde{C}^* \\
&X_{\cA' \, \cB' \, \cC} = (S_\cC\inv)^\cC - (S_{\cA'})^{\cA'} + ( S_{\cA' \cC} \; S_\cC\inv \; S_{\cC \cA'}  )^{\cA'} + (S_{\cA' \cC} \; S_\cC\inv)^{\cA' \cC} + ( S_\cC\inv \; S_{\cC \cA'})^{\cC \cA'} \, 
\eea
and a decoupled invertible $\cC$ two-form gauge theory
\be
Y_{S_\cC} = \sum_{\alpha_{\cC} \; \in \; H^2(X, \cC)} \exp\left( \frac{2 \pi i}{2 N} \fP^{S_\cC} (\alpha_\cC) \right) \, 
\ee
if $S_\cC$ is invertible. In these formulas $\Tilde{C}$ is the matrix associated to the complement of the join $\cA \smallvee \cB$. 
The formulas simplify if we gauge the full $(\bZ_N)^g$ with an invertible torsion $S$. Then $\cA' = \cB' =1$ and \eqref{eq: K-formula g} becomes
\be
\sigma(\unit) \; \tau(S) \; \sigma(\unit) = Y_{S} \; \nu(-S\inv) \; \tau(- S) \; \sigma(\unit) \; \tau(- S\inv) \, .
\ee
When the kernel of $S_\cC$ is non-empty the subgroup of $\cA \smallwedge \cB$ which is annihilated by $S_{\cA \smallwedge \cB}$ gets ``ungauged'' instead. For example suppose that $S_\cC=0$, then 
\bea
&\sigma(C_\cA) \; \tau((S_\cA)^\cA ) \sigma(C_\cB) = \nu(U_{\cA \, \cB}) \; \sigma(C_\cD \vert C_{\cB'}) \; \tau( {S_{\cA'}}^\cD ) \\
& U_{\cA \; \cB} = C_{\cD} \; C_{\cA'}^* + C_{\cB'} \; C_{\cB'}^* - C_{\cC} \; C_{\cC}^* + \Tilde{C} \; \Tilde{C}^* \, ,
\eea
where $C_\cD = (C_{\cA'} - C_\cC \; S_{\cC \cA'}) $.

\paragraph{Action on global variants.}
As in the case of $g=1$, each global variant $\cL$ can be reached from a reference boundary $\cE$ by a gauging operation $\sigma(C_\cA)$ with some discrete torsion $\tau({S_\cA}^\cA)$.
To find the explicit map we consider applying
\be
\Phi_{\cA, \, S_{\cA}} = \sigma(C_\cA) \; \tau( {S_\cA}^\cA  )  
\ee
to the electric variant $\cE = \unit_{2 g}$. Again the genuine lines after the gauging will be the gauge invariant lines in the presence of the background torsion. In our conventions Wilson lines $W_l$ are labelled by an element $l$ of the dual space. Their charge under $\cA$ gauge transformations $B \to B + C_\cA \; \delta \lambda_\cA$ is 
\be
l^\sT \; C_\cA \int \text{PD}(\gamma) \cup \lambda \, .
\ee
From this it follows that lines labelled by dual generators $C_{\Tilde{\cA}}^*$ are still genuine after the gauging. This fixes the first $g-r(\cA)$ columns of $\cL$ to be:
\be
\mat{ {C_{\Tilde{\cA}}^*}^\sT \\[-0.5em]   \\ 0 } \, .
\ee
As in the $g=1$ case, 't Hooft lines are also charged due to the discrete theta angle. The charge of $T_s$ is
\be
- s^\sT \; {C_\cA^*}^\sT (S_\cA)^\cA \int \text{PD}(\gamma) \cup \lambda \,.
\ee
Therefore neutral dyons must fulfill
\be
l^\sT \; C_\cA - s^\sT \;  {C_\cA^*}^\sT {S_\cA}^\cA = 0
\ee
which is solved by $s = C_\cA $ and $l = {C_\cA^*}^\sT \; S_\cA$. These give the last $r(\cA)$ columns of $\cL$:
\be \mat{
 {C_\cA^*}^\sT \; S_\cA \\[-0.5em] \\ C_\cA 
} \, .
\ee
It can be shown that, if $\cL = \mat{A \\ C}$, then $A^\sT \; C$ is symmetric and has rank $r(\cA)$.
\paragraph{Fusion rules.}
In a similar way as before we can also state the two basic rules for the composition in half-space:
\be
\boxed{\begin{array}{ccc}
& Y_{S_\cC} \ \text{on half space} \ X^+ \sim \ \text{decoupled TQFT-coefficient} \ \cA^{N, S_\cC} \ \text{on} \ \partial X^+ & 
\end{array}
}
\ee
and
\be
\boxed{
\begin{array}{ccc}
& \text{Un-gauging} \ \cA \subset (\bZ_N)^g \ \text{on half space} \ X^+ \sim \text{condensation defect} \ \cC^{\cA} \ \text{on} \ \partial X^+ & 
\end{array}
}
\ee
Above we have defined a generalized minimal TQFT $\cA^{N, S_\cC}$ which has line operators isomorphic to $\cC$ and spins $\theta_v = \exp\left( \frac{2 \pi i}{2 N} v^\sT S_\cC v \right)$. Since the theory is decoupled the only relevant information is contained in $S_\cC$ modulo congruence by $\GL{(r(\cC), \bZ_N)}$. When $N$ is prime one can show that there are only two inequivalent choices for each rank $r(\cC)$ (see appendix \ref{app: quadratic} for a proof of this statement), which we denote:
\be
\cN^{r, \, +} \, , \ \ \ \cN^{r, \, -} \, .
\ee
As an example, if the rank is $r=1$ there are two classes with representatives $\cA^{N,\, 1}$ and $\cA^{N, \, q' }$, where $q'$ is not a perfect square in $\bZ_N$. It is always possible to choose the representatives to be $\left( \cA^{N,1} \right)^r$ for $\cN^{r,+}$ and $\left(\cA^{N, 1}\right)^{r-1} \; \times \; \cA^{N, q'}$ for $\cN^{r, -}$.

\section{Action on line operators: the rank}\label{sec:rankQFT}
We now describe how duality defects act on line operators and introduce a new concept: the rank of a non-invertible symmetry. This can be used to almost entirely fix the fusion rules of duality defects, apart from the choice of quadratic form in the decoupled TQFT.
The action of duality defects on generic line operators for $\bZ_N$ one-form symmetry has been introduced in \cite{Choi:2021kmx}. 
The first key feature, already noted in \cite{Chang:2018iay}, is that non-invertible symmetries can lead to nontrivial maps $\fD_\cL^M : \cH_0 \to \cH_{s}$ between the untwisted ($\cH_0$) and twisted ($\cH_s$) Hilbert space. In our case the twisting is by the one-form symmetry defect $U_s$. These follow from the existence of nontrivial junctions between $\fD$ and $U_s$.
In radial quantization we represent them in the following way
\bea    
\begin{tikzpicture}     \draw[fill=black] (0,0) circle (0.1);
     \draw[color=red, line width=2] (0,0) circle (1);
     \draw[color=orange, line width=1.5] (0,1) -- (0,2);
     \draw[fill=orange] (0,1) circle (0.1);
     \node[below] at (0,0) {$W_l$};
     \node[above] at (0,2) {$U_s$};
     \node[left] at (-1,0) {$\fD_\cL^M$};
    \end{tikzpicture}
\eea
We first discuss the untwisted action. Let us suppose that $\Phi_\cL^{M\inv}$ is obtained by gauging $\cA$ with discrete torsion $S_\cA$. If $W_l$ is charged under $\cA$ then the operator is killed by the gauging interface. If it is uncharged, it will be mapped by $M$ onto another genuine operator $W_{M l}$. Consistency implies that
\begin{equation*}
\boxed{\text{Genuine line operators uncharged under $\cA$} \ = \ \text{Sublattice} \ \cK \subset \cL \ \text{ such that} \ M \cK \subset \cL }
\end{equation*}
$\cK$ can be explicitly computed as 
\be
\cK= \cL \smallwedge M\inv \; \cL \, .
\ee
Thus we conclude that
\bea
\begin{tikzpicture}
 \draw[fill=black] (0,0) circle (0.1);
     \draw[color=red, line width=2] (0,0) circle (1);
     \node[below] at (0,0) {$W_l$};
     \node[left] at (-1,0) {$\fD_\cL^M$};
     \draw[fill=black] (5.5,0) circle (0.1);
     \node[right] at (1.35,0) {$\ds = \ \ \delta_{ l \in \cK} \ \langle \fD_\cL^M(\Sigma) \rangle$};
      \node[below] at (5.5,0) {$W_{M l}$};
\end{tikzpicture}
\eea
The gauged group $\cA$, which is a subgroup $\cS_\cA \subset \cS$ must satisfy:
\be
\langle \cS_\cA , \, \cK \rangle = 0 \, .
\ee
It is clear that the ranks of the lattices satisfy $r(\cA)= g - r(\cK)$. We call $r(\cA)$ the \emph{rank} of the non-invertible defect $\fD_\cL^M$.\footnote{Sometimes we use $r(\fD)$ or $r(M)$ instead.} Duality defects with $r(\cA)=0$ are invertible. 
The maps on the twisted sector $\cH_s$ can be understood in a similar way, but now the line $W_l$ can be charged under the gauged symmetry. Consistency with the duality transformation requires that $M l = s \mod \cL $ so:
\bea
\begin{tikzpicture}
 \draw[fill=black] (0,0) circle (0.1);
     \draw[color=red, line width=2] (0,0) circle (1);
     \draw[color=orange, line width=1.5] (0,1) -- (0,2);
     \draw[fill=orange] (0,1) circle (0.1);
     \node[below] at (0,0) {$W_l$};
     \node[left] at (-1,0) {$\fD_\cL^M$};
      \node[above] at (0,2) {$U_s$};
            \draw[color=orange, line width=1.5] (5.5,0) -- (5.5,2);
     \draw[fill=black] (5.5,0) circle (0.1);
     \node[right] at (1.35,0) {$ = \ \ \ds \delta_{ [M l] , \, s} \ \langle \fD_\cL^M(\Sigma) \rangle$};
      \node[below] at (5.5,0) {$T_{M l}$};
        \node[above] at (5.5,2) {$U_s$};
\end{tikzpicture}
\eea
The characterization of the rank of the symmetry using $\cK$ can be used to understand the structure of the fusion algebra without performing the direct computation. First notice that the rank can be written as
\bea
\begin{tikzpicture}
\node[left] at (0,0) {$\ds r(\fD_\cL^M) = \ g \ - \ $};
 \draw (0,-0.75) -- (0,0.75) ;
\draw[color=red] (0.75,0) circle (0.7);
\draw[fill=black] (0.75,0) circle (0.05);
\node at (0.75,-0.9) {\scriptsize $ M$};
\draw (1.5,-0.75) -- (1.5,0.75) ;
\end{tikzpicture}
\eea
where $\vert \cdot \vert$ is the dimension of the image of the map inside. After computing the fusion we define
\bea \label{eq: rank21}
\begin{tikzpicture}
\node[left] at (0,0) {$\ds r\left(\fD_\cL^{M_2} \times \fD_\cL^{M_1} \right) = g \ - \  $};
\draw (0,-0.75) -- (0,0.75) ;
\draw[color=red] (0.75,0) circle (0.7);
\draw[color=red] (0.75,0) circle (0.3);
\draw[fill=black] (0.75,0) circle (0.05);
\node at (0.75,-0.5) {\scriptsize $ M_1$};
\node at (0.75,-0.9) {\scriptsize $ M_2$};
\draw (1.5,-0.75) -- (1.5,0.75) ;
\end{tikzpicture}
\eea
Notice that this is different from
\bea \label{eq: rank21fus}
\begin{tikzpicture}
\node[left] at (0,0) {$\ds r(\fD_\cL^{M_2 M_1}) = \ g \ - \ $};
 \draw (0,-0.75) -- (0,0.75) ;
\draw[color=red] (0.75,0) circle (0.7);
\draw[fill=black] (0.75,0) circle (0.05);
\node at (0.75,-0.9) {\scriptsize $ M_2  M_1$};
\draw (1.5,-0.75) -- (1.5,0.75) ;
\end{tikzpicture}
\eea
because the image of \eqref{eq: rank21} is spanned by $\cK_{M_1} \smallwedge M_2\inv \cK_{M_1} $ while for \eqref{eq: rank21fus} by $\cK_{M_{2,1}}$ which is a larger vector space. When the two do not agree a further object is needed to make the fusion consistent. This is a condensate $\cC_{2,1}$. It's rank is computed by
\bea
\begin{tikzpicture}
\node[left] at (0,0) {$\ds r\left(\cC_{2,1}\right) = $};
 \draw (0,-0.75) -- (0,0.75) ;
\draw[color=red] (0.75,0) circle (0.7);
\draw[fill=black] (0.75,0) circle (0.05);
\node at (0.75,-0.9) {\scriptsize $ M_2  M_1$};
\draw (1.5,-0.75) -- (1.5,0.75) ;
\node at (2.25,0) {$-$};
\begin{scope}[shift={(3,0)}]
 \draw (0,-0.75) -- (0,0.75) ;
\draw[color=red] (0.75,0) circle (0.7);
\draw[color=red] (0.75,0) circle (0.3);
\draw[fill=black] (0.75,0) circle (0.05);
\node at (0.75,-0.5) {\scriptsize $ M_1$};
\node at (0.75,-0.9) {\scriptsize $ M_2$};
\draw (1.5,-0.75) -- (1.5,0.75) ;
\end{scope}
\end{tikzpicture}
\eea
This information is readily obtainable as soon as we know $\cK$ for the various defects.

\section{The $5d$ Symmetry TFT}
\label{sec:The $5d$ Symmetry TFT}
Another viable path to compute the fusion algebra of the duality symmetries is using the bulk Symmetry TFT description \cite{Antinucci:2022vyk}. 
The symmetry TFT is a $d+1$-dimensional topological theory which encodes the discrete symmetries of a given $d$-dimensional QFT. More specifically a $d$-dimensional \emph{absolute} QFT is isomorphic to a \emph{relative} QFT living at the boundary of a $(d+1)$-dimensional slab where the symmetry TFT lives. At the other boundary one should impose gapped (topological) boundary conditions $\cL$ which select the given global structure of the absolute theory (see figure \ref{fig: freedsandiwch}). 

In our applications the boundary conditions are of Dirichlet type for the generators of the Lagrangian algebra $\cL$\footnote{In this sense $\rho$ is equivalent to perform a gauging of a Lagrangian algebra on the bulk TFT\cite{Kaidi:2021gbs,Benini:2022hzx}.}. 
Defects in the quotient $\cS = \Gamma/\cL$, when pushed to the boundary, are the topological operators of the $d$-dimensional absolute theory. Operators in $\cL$ can instead end the topological boundary, and their endpoints describe operators charged under $\cS$. 

\bea
\begin{tikzpicture}[scale=1.5]		\node at (2.5,1.75) { \textbf{Symmetry}};
			\filldraw[fill=white!70!blue, opacity=0.5] (0,0) -- (1,0.25) -- (1,1.25) -- (0,1) -- cycle;
	\draw[color=blue, line width=1] (0,0.5) to[bend left=30] (0.5,0.625); 	\draw[color=blue, line width=1] (0.5,0.625) to[bend right=30] (1,0.75);
	\node[above] at (0.5,0.625) {$\cS$};
		\node at (1.5,0.625) {$=$};
		\begin{scope}[shift={(2,0)}]
			\draw (0,0) --(1.5,0); \draw (1,0.25) -- (2.5,0.25); \draw (1,1.25) -- (2.5,1.25); \draw (0,1) -- (1.5,1);
			\filldraw[fill=white!70!blue, opacity=0.5] (1.5,0) -- (2.5,0.25) -- (2.5,1.25) -- (1.5,1) -- cycle;	
				\draw[color=red, line width=1] (0,0.5) to[bend left=30] (0.5,0.625); 	\draw[color=red, line width=1] (0.5,0.625) to[bend right=30] (1,0.75);
					\node[above] at (0.5,0.625) {$\cS$};
			\filldraw[fill=white!70!red, opacity=0.5] (0,0) -- (1,0.25) -- (1,1.25) -- (0,1) -- cycle;
			\end{scope}	

	 	 \begin{scope}[shift={(5,0)}]
	 	\node at (2.5,1.75) { \textbf{Charged operators}};
	 	\filldraw[fill=white!70!blue, opacity=0.5] (0,0) -- (1,0.25) -- (1,1.25) -- (0,1) -- cycle;
	 	\draw[fill=olive] (0.5,0.625) circle (0.05);
	 		\node[above] at (0.5,0.625) {$\cW$};
	 	\node at (1.5,0.625) {$=$};
	 	\begin{scope}[shift={(2,0)}]
	 			\filldraw[fill=white!70!red, opacity=0.5] (0,0) -- (1,0.25) -- (1,1.25) -- (0,1) -- cycle;	
	 			\draw[fill=olive] (0.5,0.625) circle (0.05);
	 			\draw[color=green, line width=1] (0.5,0.625) -- (2,0.625); 
	 				\node[above] at (1.25,0.625) {$\cW$};
	 		\draw (0,0) --(1.5,0); \draw (1,0.25) -- (2.5,0.25); \draw (1,1.25) -- (2.5,1.25); \draw (0,1) -- (1.5,1);
	 		\filldraw[fill=white!70!blue, opacity=0.5] (1.5,0) -- (2.5,0.25) -- (2.5,1.25) -- (1.5,1) -- cycle;	
	 		\draw[fill=olive] (2,0.625) circle (0.05);
	 	\end{scope}
	 \end{scope}
	\end{tikzpicture}
\eea
Recently, various authors have shown in detail how to derive the Symmetry TFT for theories with an holographic dual starting from string/M-theory \cite{Apruzzi:2021nmk, Bashmakov:2022jtl, Apruzzi:2022rei, Antinucci:2022vyk, vanBeest:2022fss, Bashmakov:2022uek,Bah:2020uev}. The symmetry TFT for the duality defects in $\cN = 4$ SYM was first introduced in \cite{Kaidi:2022cpf}, while in \cite{Antinucci:2022vyk} it was embedded in a holographic framework and the fusion rules for such defects are derived from the bulk formalism. In this section we extend this analysis to the case $g>1$ and explain how features introduced in Section \ref{sec:Duality defects and discrete topological manipulations} emerge form the bulk.

The symmetry TFT for 6d $\cN=(2,0)$ theories of type $A_{N-1}$ has a simple holographic derivation, since the theory can be realized by a stack of $N$ M5 branes in flat space-time.
These induce $N$ units of $G_4=d\cC$ flux on a round $S^4$, where $\cC$ is the 3-form potential. The symmetry TFT can be derived from the reduction of the topological terms in the eleven dimensional supergravity action. The resulting $7$-dimensional topological theory is a Chern-Simons theory \cite{Monnier:2017klz}
\be\label{eq: 7d CS}
S_{7d} = \frac{N}{4\pi} \int_{Y_7}  \cC \, d \cC \, .
\ee
The operators of the theory are 
\be
C_m = e^{i m \int \cC} \ \ , \ \ \ \ m=0,...,N-1
\ee
generating a $\bZ_N^{[3]}$ 3-form symmetry. The TFT \eqref{eq: 7d CS} generically does not have gapped boundary conditions, indicating the fact that $6$d $\cN = (2,0)$ SCFTs are intrinsically relative \cite{Witten:2009at}.

The symmetry TFT for class $\cS$ theories is obtained by considering a seven-dimensional space-time of the from $X_7 = X_5 \times \Sigma_g$ and reducing on $\Sigma_{g}$. The $5$d symmetry TFT takes the form\footnote{We assume that $H_2(X_5, \bZ)$ is trivial.}
\begin{equation}\label{eq:classSsymmTFT}
    S_{5d} = \frac{N}{4\pi} \int \sum\limits_{i,j=1}^{2g}\cB_i  \; \cJ_{ij} \;  d\cB_j \ , \ \ \  \qquad \cJ = \mat{0 & \unit_g \\ -\unit_g & 0}.
\end{equation}
However, this is too naive. In the eleven-dimensional theory, all the global symmetries are gauged since gravity is not decoupled. Whenever we choose a vacuum in such gravity theories, part of these gauge symmetries is Higgsed and appears as global symmetries of the IR effective theory. Among these are the large diffeomorphisms of $\Sigma_g$, for which the order parameter can be taken as the complex structure matrix $\Omega$. However, if our choice of $\Sigma_g$ has discrete isometries, we will obtain a residual gauge symmetry in the compactified theory. Thus, whenever the Riemann surface has a nontrivial automorphism group $G(\Omega) \subset \Sp (2g, \bZ)$, we obtain an emergent gauge symmetry in the symmetry TFT. This acts on charge labels of the 2-form gauge fields, transforming in the fundamental representation of $\Sp(2g, \bZ_N)$. We will argue that this emergent gauge symmetry is responsible for the non-invertible duality defects

An analogous scenario is enjoyed by $\cN=4$ SYM when viewed as the theory on D3 branes in type IIB string theory, at $\tau=i$ or $e^{2\pi i/3}$. From a holographic point of view, the axio-dilaton VEV of type IIB string theory Higgses the $\SL(2,\bZ)$ gauge symmetry, which consequently appears as a global symmetry of the supergravity theory. However, at $\tau=i$ or $e^{2\pi i/3}$, the VEV of such a field is invariant under the $\bZ_{4,6}$ subgroup of $\SL(2,\bZ)$. Consequently, an emergent gauge field remains present in the infrared. In \cite{Antinucci:2022vyk}, it was shown that this gauge field is responsible for the topological duality (and triality) defects of the dual gauge theory. The same conclusion is reached if we regard $\cN=4$ SYM as the theory obtained by compactifying the 6d $\cN=(2,0)$ theory on a torus with an appropriate modular parameter, $\tau=i$ or $e^{2\pi i/3}$. Indeed, M-theory compactified on a small torus is equivalent to type IIB string theory \cite{Johnson:1997wf}, in which S-duality is realized as a modular transformation of the torus. Thus, we see that from the M-theory perspective, the maximally supersymmetric case analyzed in \cite{Antinucci:2022vyk} is not special, and the un-Higgsed subgroup can be embedded in the group of large diffeomorphisms in string theory.

\subsection{Duality defects from the Symmetry TFT}
Let us describe the construction of duality defects in the Symmetry TFT and how to compute fusion rules. Most of the results are a straightforward generalization of \cite{Antinucci:2022vyk}, to which we refer for a thorough analysis. 
The symmetries of \eqref{eq:classSsymmTFT} are a $2$-form symmetry $(\bZ_N^{[2]})^{2g}$ generated by the topological surface operators
\be
U_m = e^{im^{\sT}\int \cB}
\ee
and a zero-form symmetry $\Sp (2g,\bZ _N)$ acting on the gauge fields as
\be
\cB \rightarrow M^\sT \cB \ , \ \ \qquad M \in \Sp(2g,\bZ_N) \ .
\ee
All the topological defects implementing this zero-form symmetry are condensation defects \cite{Roumpedakis:2022aik} constructed by higher-gauging a subgroup $\cA$ of the 2-form symmetry on a codimension one manifold with an appropriate choice of discrete torsion.

Given a subgroup $\cA\subset (\bZ_{N})^{2g}$ and a symmetric torsion matrix $\cT_{\cA}$ we can define a condensation defect on a compact four manifold $\Sigma$ with $H_2(\Sigma,\bZ) = \bZ^2$ as a sum
\be\label{eq:condensationSp}
V[\cT_\cA] = \sum_{m,m' \in \cA} \exp\left( -\frac{2 \pi i}{N} \left( m^\sT \left( \cT_{\cA} + 2^{-1} \cJ_\cA \right) m'  \right) \right) \; U_{C_\cA m'}(\gamma) \; U_{C_\cA m}(\gamma') \, ,
\ee
with $\gamma$ and $\gamma'$ being the generators of $H_2(\Sigma,\bZ)$.\footnote{We will follow the notation of section \ref{sec:Duality defects and discrete topological manipulations} for the lattice operations.}
To construct a topological defect which implements $M\in \Sp{(2g,\bZ_N)}$ we need to impose the correct group action on surface operators $U_k$. Surrounding $U_k(\gamma)$ by the defect $V[\cT_\cA]$ we have
\bea
V[\cT_\cA]   U_k(\gamma)  =& \sum_{m,m' \in \cA} \exp\left( -\frac{2 \pi i}{N} \left( m^\sT \left( \cT_{\cA} + 2^{-1}  \cJ_\cA  \right) m'  + m^\sT C_\cA^* \cJ k \right)\right) U_{k+C_\cA m'}(\gamma) \,  \\
=& \, U_{M \cdot k}(\gamma) \, ,
\eea
where we have shrunk $V[\cT_\cA]$ at end. This implies
\be
M = \left( \unit _{2g} - C_\cA \left( \cT_{\cA} + 2^{-1} \cJ_\cA \right)^{-1} C_\cA^* \cJ \right) \, . \label{eq: MfromT}
\ee
 Notice that the image of $(M -\unit_{2g})$ is isomorphic to $\cA$. Inverting \eqref{eq: MfromT} we find\footnote{Let us consider for example
\be
M = \mat{\unit_g & B\\ 0&\unit_g}
\ee
with $B$ non-singular. The image of $M-\unit_{2g}$ is contained in the ``electric'' $(\bZ_{N})^{g}$. To implement this symmetry it suffices to gauge the aforementioned electric $(\bZ_N)^g$ only.}
\be
\cT_\cA = - 2^{-1} \cJ_\cA + \left[\left((M - \unit_{2g}) \cJ\right)_\cA  \right]^{-1}\ 
\ee
This determines univocally the zero-form symmetry defects.
For a full gauging $C_\cA = \unit_{2g}$ and we get
\be
\label{eq:T(M)}
\cT = 2^{-1} \cJ \left( \unit_{2g} + M \right)\left( \unit_{2g} - M\right)^{-1} \, ,
\ee
which generalizes \cite{Antinucci:2022vyk}.

The fusion of two condensation defects can be computed from \eqref{eq:condensationSp}. After some algebra we find that
\be
V[\cT^{(1)}_{\cA}] \times V[\cT^{(2)}_{\cB}]  = V[\cT^{(2,1)}_{\cA,\cB}]\, ,
\ee
where
\be
\cT^{(2,1)}_{\cA, \cB} = \cT^{(2)}_{\cA\smallvee\cB} -  \left(\cT^{(2)}_{\cA\smallvee\cB, \cA} - 2^{-1}\cJ_{\cA \smallvee \cB, \cA}\right)\left(\cT_{\cA}^{(1)}+ \cT_{\cA}^{(2)}\right)^{-1}\left(\cT^{(2)}_{\cA\smallvee\cB, \cA} + 2^{-1}\cJ_{\cA \smallvee \cB, \cA}\right)\,
\ee
is the torsion matrix corresponding to the $\Sp(2g,\bZ_N)$ element $M_2M_1$. Following \cite{Antinucci:2022vyk} we can give a Lagrangian description of the defect
\be\label{eq: contiuous acton 4d}
S_V = \frac{N}{4\pi}\int_{\Sigma_4} \cB_\cA ^\sT \Phi_\cA + \Gamma_\cA^\sT d \cB_\cA + \Psi_\cA^\sT d\Phi_\cA +\frac{1}{2}\Phi_\cA^\sT \cT_\cA \Phi_\cA
\ee
where $\Phi_\cA,\Gamma_\cA,\Psi_\cA$ are auxiliary fields.\footnote{Alternatively, in discrete notation it can be written as 
\be \frac{2\pi}{N} \int \cB_\cA \cup \Phi_\cA + \frac{1}{2} \fP^{T_\cA}(\Phi_\cA) \, ,
\ee
with $\Phi_\cA \; \in \; H^2(\Sigma_4, \; \cA)$ and $B_\cA \; \in \; H^2(\Sigma_4, \; \cA^*)$.
} On a closed $\Sigma_4$ this is gauge invariant under
\be
\begin{split}
    & \cB_\cA \to \cB_\cA + d \alpha_\cA \, , \ \ \Phi_\cA \to \Phi_\cA + d \lambda_\cA \,, \\
   & \Psi_\cA \to \Psi_\cA - \cT_\cA \lambda_\cA - \alpha_\cA + d \mu_\cA \, , \  \ \Gamma_\cA \to \Gamma_\cA - \lambda_\cA + d \nu_\cA  \, .
\end{split}
\ee

\paragraph{Twisted sectors and fusion rules.}
We can now define the twist defects associated to zero-form symmetry operators $V[\cT_\cA]$. Given a $p$-dimensional topological operator $V$, we can consider its twisted Hilbert space $\cH_V$, which is spanned by non-genuine $p-1$ dimensional topological defects on which $V$ can end. If $V$ implements an anomaly-free symmetry, gauging $V$ liberates the twist defects, which become genuine $p-1$ dimensional operators in the gauged theory. 
To construct the twist defects $D[\cT_\cA]$ for $V[\cT_\cA]$ we can impose Dirichlet boundary conditions on $\Phi_\cA$.\footnote{This is only true if $T_\cA$ is an invertible matrix, in which case the defect theory is an invertible TQFT which has only one allowed boundary condition. Studying twist defects for which $T_\cA$ is not full rank is challenging and we do not consider them in this work.} The minimal description of $D[\cT_\cA]$ can be found by compensating for the lack of gauge invariance on an open $\Sigma_4$ by the following action on $Y=\partial \Sigma_4$:\footnote{In discrete notation this reads \be
-\frac{2 \pi}{2} \int \gamma_\cA \; \cup \cT_\cA \; \beta(\gamma_\cA) \, , \ \ \ \gamma_\cA \; \in \; H^1(Y, \cA) \, .
\ee
with $\beta$ the Bockstein map.}
\be\label{eq: action twist}
S_{twist} = \frac{N}{4\pi} \int_Y \cB_\cA^\sT \Gamma_\cA + \Phi_\cA^\sT \Psi_\cA + \Gamma_\cA d \Psi_\cA - \frac{1}{2}\Gamma_\cA^\sT \cT_\cA d\Gamma_\cA \ .
\ee

If $\cT_\cA$ is full rank we can obtain a simpler description of $D[\cT_\cA]$ by integrating out $\Phi_\cA$. The 4d defect $V[\cT_\cA]$ becomes\footnote{By integrating out $\Gamma _\cA$ one can rewrite this expression in the discrete formalism as 
\begin{equation}
- \frac{2\pi i}{2 N} \int \fP^{\cT\inv_\cA} (\cB_\cA)
\end{equation}
, where now $\cB_\cA$ is understood to be a discrete gauge field.}
\be\label{eq: 4d anomaly}
S_V = \frac{N}{2\pi}\int_{\Sigma_4} \cB^\sT_\cA d \Gamma_\cA - \frac{1}{2}\cB^\sT_\cA \cT\inv_\cA \cB_\cA  
\ee
The corresponding twist defect is described by a minimal $\cA^{N, -\cT_\cA}$ TQFT \cite{Hsin:2018vcg}. This is a 3d TQFT hosting $N^{r(\cA)}$ lines $W_n$, which fuses according to $\cA$ and have spins  
\be
\theta(W_n) = \exp{\left(\frac{\pi i}{N} n^{\sT}\cT_\cA n\right)}\,.
\ee
A simple way to derive this fact is to interpret \eqref{eq: 4d anomaly} as an anomaly-inflow action for a $\cA$ one-form symmetry in 3d. The minimal TQFT $\cA^{N, -\cT_\cA}$ is the minimal\footnote{By minimal we mean that each theory $\cT$ with the same anomaly can be written as $\cT = \cA^{N, \cT_\cA} \times \cT'$ for some $\cT'$ decoupled from $\cA$.} possible MTC saturating the anomaly \cite{Hsin:2018vcg}. If $\cT_\cA$ is not full-rank this reasoning fails as the 4d theory is not an invertible TQFT and to use such a description one should presumably make a choice of boundary condition on $\partial \Sigma_4$ first. We do not discuss such cases here.
Notice that if $\cA \neq (\bZ_N)^{2g}$ the twist defect is not unique as we can always fuse it to a 3d condensate of $(\bZ_N)^{2g}/\cA$.

Fusion rules for twist defects $D[\cT^{(1)}_\cA]$ and $D[\cT^{(2)}_\cB]$ can be computed by noticing that lines $W^{(1)}$ and $W^{(2)}$ are not mutually local, as they are attached to bulk $U_m$ surfaces \cite{Antinucci:2022vyk}. Taking this into account, lines of the composite defect $D[\cT^{(1)}_\cA] \times_\cB D[\cT^{(2)}_\cB]$ (where $\times_\cB$ reinforces that it is not a naive tensor product) braid through the braiding matrix
\be
\cK_{21} = \mat{\cT^{(1)}_\cA & 2\inv \; \cJ_{\cA,\cB}\\ - 2\inv \;\cJ_{\cB,\cA} & \cT^{(2)}_\cB} \,.
\ee
To compute the fusion product one must isolate the group of uncharged lines under the bulk 1-form symmetry. The remaining coupled theory is precisely the twist defect $D[\cT^{(1,2}]$.
For instance, when $C_\cA = C_\cB = \unit$, we find
\be
D[\cT^{(1)}]\; \times \; D[\cT^{(2)}] = \cA^{N,-\cT^{(1)}-\cT^{(2)}} \times D[\cT^{(2,1)}] \, ,
\ee
as long as $\cT^{(2,1)}$ also has full rank. In the opposite case, when $\cT^{(2)} = - \cT^{(1)}$ and the final zero-form symmetry defect is the identity, we find
\be
D[\cT^{(1)}]\; \times \; D[\cT^{(2)}] = \cC^{(\bZ_N)^{2g}} \, .
\ee
\paragraph{Fusion on gapped boundaries.}
What we are actually interested in is to discuss the composition laws for twist defects once they are brought onto a gapped boundary $\cL$.
Before the gauging of the zero-form symmetry $G(\Omega)$, these describe fusion rules for duality \emph{interfaces}
\bea
    \begin{tikzpicture}[scale=2.5]
				 \draw (1,0.25) -- (3,0.25); \draw (1,1.25) -- (3,1.25);

			\filldraw[fill=white!70!red, opacity=0.5] (0,0) -- (1,0.25) -- (1,1.25) -- (0,1) -- cycle;	

  \filldraw[fill=white!70!olive, opacity=0.5] (2.75,0.1925) -- (2.75, 1.1925 ) -- (0.75, 1.1925) -- (0.75,0.1925) -- cycle;

  	\draw[color=olive, line width=1.5] (0.75,0.1925) -- (0.75, 1.1925);

  \filldraw[fill=white!70!green, opacity=0.5] (2.25,0.0675) -- (2.25, 1.0675 ) -- (0.25, 1.0675) -- (0.25,0.0675) -- cycle;
    
	  	\draw[color=green, line width=1.5] (0.25,0.0625) -- (0.25, 1.0625);
		
   \filldraw[fill=white!70!blue, opacity=0.5] (2,0) -- (3,0.25) -- (3,1.25) -- (2,1) -- cycle;	
    \draw[color=black, line width=1] (2.25,0.0675) -- (2.25, 1.0675 );
  
      \draw[color=black, line width=1] (2.75,0.1925) -- (2.75, 1.1925 );
      \draw (0,1) -- (2,1); \draw (0,0) --(2,0);

      \node[above] at (0.1,1.1) {$D[\cT^{(1)}]$};
       \node[above] at (0.7,1.25) {$D[\cT^{(2)}]$};
\end{tikzpicture}
\eea
Here we will mostly discuss the case in which $\cA = \cB = (\bZ_N)^{2g}$. However we give a general algorithm at the end of the section.
In section \ref{sec: global variants} we have learned that global variants correspond to Lagrangian lattices $\cL$ of $(\bZ_N)^{2g}$. From the symmetry TFT perspective this is encoded in a boundary condition which sets:
\be
U_m = \unit \, , \ \ \ \text{if} \ m \in \cL
\ee
or, alternatively on the fields
\be\label{eq: boundary conditions}
\cB_{\cL}\equiv \cL^{\sT}\cB = 0 \qquad (\text{up to gauge transformations}).
\ee
There are two cases to discuss, depending on whether the rank of $\cT_\cL$ is maximal or not. Let us start by assuming that $\cT_\cL$ is an invertible $g \times g$ matrix. We will treat the other cases in \ref{sec:rank bulk}.

Since on the gapped boundary a subgroup $\cL$ of the $(\bZ_N)^{2g}$ two-form symmetry acts trivially, lines $W_l$ with $ l \in \cL$ of a twist defect $D[\cT]$ which are only charged under $\cL$ completely decouple. These form a $\cA^{N, - \cT_\cL}$ minimal theory, which we consider screened on the boundary.\footnote{Notice that this is a well defined MTC only if $\cT_\cL$ is invertible.} The coupled twist defect is described by:
\be
D_\cL[\cT] = \frac{D[\cT] \times \cA^{N, T_\cL}}{ \cL } \, .
\ee
If $\cT$ is full rank, we can faithfully parametrize lines surviving the quotient by $\cS = \cT\inv \cL_\perp$, with $\cL_\perp = \cJ \; \cL$ the dual lattice of $\cL$. These lines form a minimal theory 
\be
\cA^{N, - \cT_\perp} \, , \ \ \text{with} \ \cT_\perp = \cT\inv_{\cL_\perp} \, .
\ee
A complementary procedure, which leads to the same answer, is to impose the boundary conditions $\cL^\sT \cB = 0$ directly on the ``anomaly inflow'' action \eqref{eq: 4d anomaly}. The leftover anomaly is precisely captured by the minimal theory $\cA^{N, -\cT_\perp}$.

Boundary fusion rules can be computed by applying the screening process to the $\cA^{N,\cT^{(1)}}\times_{\cB}\cA^{N,\cT^{(2)}}$ theory instead. We define:
\be
\cA^{N, \cR_{2,1}} = \frac{  \cA^{N,\cT^{(1)}}\times_{\cB}\cA^{N,\cT^{(2)}}  \times  \cA^{N, \cT^{(1)}_\cL}   \times \cA^{N, \cT^{(2)}_\cL} } {\cL \times \cL} \ .
\ee
When $\cT^{(2,1)}$ and its restriction to $\cL$ are also full rank we can parametrize 
\be
  \cR_{21} = \mat{\cT^{(2,1)}_{\perp} & c_0 \\c_0 & c_d} \ \ \ \ \ \ \ 
\begin{aligned}
&c_0 = \cL_{\perp}^{\sT}\;{\cT^{(2,1)}}^{-1} \; \left(\cT^{(2)}-2\inv\;\cJ\right)\left(\cT^{(1)}+\cT^{(2)}\right)^{-1} \;\cL_{\perp} \\
 & c_d = \cL_{\perp}^{\sT} \; \left(\cT^{(1)} + 4\inv \cJ\; {\cT^{(2)}}^{-1} \; \cJ \right) \; \cL_{\perp} \; .
\end{aligned}
\ee
This theory splits into the outgoing defect $D_\cL[\cT^{(2,1)}]$, described by the upper-left corner of the matrix, and a decoupled TQFT coefficient which can be computed on a case by case basis. The fusion rules then read
\be
D_\cL[\cT^{(1)}] \; \times \;  D_\cL[\cT^{(2)}] =  \cN_{21}\; D_\cL[\cT^{(2,1)}] \, .
\ee
If the rank of $\cT^{(2,1)}_\cL$ decreases instead, the fusion is accompanied by a condensation. The next part of this section will clarify how to treat this case.

\subsection{Lower rank defects: rank from the bulk} \label{sec:rank bulk}
We now discuss the case in which the matrix $\cT_\cL$ has a kernel. It is a matter of simple algebra to show that $\cT_\cL$ has a kernel if and only if the sublattice $\cK = \cL \; \smallwedge \; M\inv \cL$ which is mapped inside $\cL$ by $M$ is non empty. 
The kernel is then spanned by vectors
\be
\cK_{\text{Ker}} = (\unit _{2g} - M) \; \cK \, .
\ee
Let's now discuss the implications of a nontrivial $\cK_{\text{Ker}}$ on the boundary defect $D_\cL(\cT)$. It is clear that now our previous algorithm to screen decoupled lines fails, as the $\cA^{N, - \cT_\cL}$ theory is ill defined. Instead lines in $\cK_{\text{Ker}}$ form a condensable subgroup of $\cA^{N, \cT}$ and are gauged by the boundary conditions. Defining $\cL_\cK = \cL / \cK_{Ker}$ the correct prescription is
\be
D_\cL[\cT] = \frac{\cA^{N, - \cT} \times \cA^{N, \cT_{\cL_\cK}}}{\cL} \label{eq: prescription restriction}
\ee
where we have used that $\cL_\cK \times \cK_{\text{Ker}} = \cL$. The same conclusion can be reached also from the anomaly theory \eqref{eq: 4d anomaly}: when $\cK$ is non-empty the anomaly theory is of lower rank, with a kernel spanned by $\cJ \; (M + \unit _{2g}) \; \cK$.
In both cases the rank of the boundary defect decreases by $\dim(\cK)$, which reproduces the result found in Section \ref{sec:rankQFT}. 

We can now state an algorithm to compute fusion rules from the bulk Symmetry TFT.
\begin{enumerate}
    \item Find the groups $\cK_1$, $\cK_2$ and $\cK_{2,1}$ of the ingoing and outgoing defects.
    \item Construct the boundary theory for the fusion
    \be
\cA^{N, \cR_{2,1}} = \frac{  \cA^{N,\cT^{(1)}}\times_{\cB}\cA^{N,\cT^{(2)}}  \times  \cA^{N, \cT^{(1)}_{\cL_\cK}}   \times \cA^{N, \cT^{(2)}_{\cL_\cK}} } {\cL \times \cL} \, .
    \ee
\item Find a splitting:
\be
\cA^{N, \cR_{2,1}} = \cN_{2,1} \; \times \cC^{\cA_{\text{Cond}}} \; \times D_\cL[\cT^{(2,1)}]\,.
\ee
This is tedious but doable, since the condensed lines are neutral under the symmetry of $D_\cL[\cT^{(2,1)}]$ .
\end{enumerate}

\paragraph{Rank and Fusion.} Having explained how the rank of a duality defect can be understood both from a 4d and 5d perspective, we conclude by giving an explicit application of this concept.

It is rather simple, given defects $\fD_\cL^{M_1}, \, \fD_\cL^{M_2}$ and $\fD_\cL^{M_{1} M_2}$, to compute the associated groups $\cK_1$, $\cK_2$ and $\cK_{1,2}$. We now argue that this information almost unequivocally fixes the fusion algebra $\fD_\cL^{M_1} \times \fD_\cL^{M_2}$. 
First notice that the composite defect $\fD^{M_1}_\cL \times \fD^{M_2}_\cL$ host $N^{r(\fD^{M_1}_\cL) + r(\fD^{M_2}_\cL)  }$ non-genuine line excitations, on which the gauged one-form symmetry surfaces can end:
\bea
\begin{tikzpicture}
    \filldraw[ color=white, fill=white!70!red] (0,0) -- (2,-0.5) -- (2,1.5) -- (0,2) -- cycle;
    \draw[line width=1] (1,0.75) circle (0.3);
    \draw[fill=white!50!blue, opacity=0.5] (1,1.05) arc (90:-90:0.3 and 0.3) -- (-1,0.45) -- (-1,1.05) -- cycle;
    \draw[fill=white!90!blue, dashed] (-1,0.75) circle (0.3);
    \node[below] at (1,-0.5) {$\fD_\cL^M$};
    \node[left] at (-1.3, 0.75) {$U_m$};
    \node[below] at (1,0.45) {$L_m$};
\end{tikzpicture}    
\eea
These can be thought of as two-morphisms $L_m : \unit_{\fD_\cL^M} \times U_m \to \unit_{\fD_\cL^M}$ and their category is explicitly described in the bulk symmetry TFT by a minimal TQFT \eqref{eq: prescription restriction}.

The number of these lines cannot change upon performing fusion, so it must be matched on the other side. We have $N^{r(\fD_\cL^{M_2 M_1})}$ lines from the outgoing defect and $N^{2 r\left(\cC_{2,1}\right) }$ lines from the condensation defect. The factor of 2 comes from regarding the condensate as a DW theory coupled to a dynamical two-form field. The subgroup of the one-form symmetry which is condensed can be computed as a quotient $\cA_{\text{Cond}} = \left( \cA_1 \smallvee \cA_2 \right) / \cA_{1,2}$. 
The remaining lines will necessarily form a decoupled TQFT. This determines the rank of the TQFT coefficient. 
The only undetermined datum is its class modulo congruence.

\subsection{The gauged theory}
The correct 5d bulk theory describing the special loci of the conformal manifold where we get the duality defects is obtained from \eqref{eq:classSsymmTFT} by gauging the automorphism group $G(\Omega) \subset \Sp (2g,\bZ _N)$ of the Riemann surface on which we compactify the 6d $\cN=(2,0) $ theory and the 7d TQFT. We will henceforth drop the suffix $\Omega$.

The gauging consists in coupling \eqref{eq:classSsymmTFT} to a pure $G(\Omega)$ gauge theory, which renders the four-dimensional defects labelled by $M\in G(\Omega)$ transparent.
Their twist defects become genuine three-dimensional operators, provided we dress them by the naive Gukov-Witten operators (GW) of the pure $G(\Omega)$ gauge theory, as explained in \cite{Antinucci:2022vyk}. 
A convenient way to describe the gauged theory is through the \emph{hybrid formulation} introduced in \cite{Antinucci:2022vyk}, in which the bulk 2-form fields $\cB _i$ are kept continuous, while the 1-form $G$ gauge field $A$ is a singular (or Cech) cochain. 
The general treatment goes in parallel to the case of $\cN=4$ SYM (namely $g=1$) studied in \cite{Antinucci:2022vyk}. Here we focus on the main differences when $G(\Omega)$ is non-abelian. 

The first issue is defining non-abelian discrete gauge fields, which are expected to be described by $H^1(X_5,G)$.
However this object is not a group and its definition is somewhat subtle. Let us sketch how this is done (see e.g. chapter 7 of \cite{Wedhorn}).

We choose a good cover $X_5=\bigcup _i \cU _i$ of the manifold, and we assume an ordering for the indices. The covering is dual to a simplicial decomposition: patches are associated with vertices (or 0-simplices) $v_i\in \cU_i$, double intersections $\cU_{ij}$ to 1-simplices $v_{ij}$, $i<j$, crossing a co-dimension one plane and oriented from $v_i$ to $v_j$, and triple intersections $\cU_{ijk}$ are associated with co-dimension two planes orthogonal to 2-simplices $v_{ijk}$, with $i<j<k$. 
A zero cochain $\lambda \in C^0(X,G)$ associates an element $\lambda_i\in G$ to each vertex $v_i$, while a 1-cochain $A\in C^1(X,G)$ is an assignment of $A_{ij}\in G$ for each 1-simplex $v_{ij}$, and so on. Given $A,B\in C^p(X,G)$, the group structure of $G$ is used to construct $AB \in C^p(X,G)$
\begin{equation}
    (AB)_{i_0,...,i_p}=A_{i_0,...,i_p}B_{i_0,...,i_p} \in G \ .
\end{equation}
making $C^p(X,G)$ a group. We would like to define differentials $\delta _p : C^p(X,G)\rightarrow C^{p+1}(X,G)$ such that  $\delta _{p+1}\delta _p=0$. For generic $p$ this is not possible, but fortunately we only need the cases $p=0,1$, for which we introduce
 \begin{equation}
        (\delta _0\lambda)_{ij}=\lambda _i \lambda _j^{-1} \ \ \ \ \ \ \ (\delta _1 A)_{ijk}=A_{jk}A_{ik}^{-1}A_{ij} \ ,
    \end{equation}
satisfying $\delta _1\delta _0=0$. These maps are \emph{not} homomorphisms and thus $\text{Ker}(\delta _1)$ and $\text{Im}(\delta _0)$ are not groups. However on $\text{Ker}(\delta _1)$ we can introduce the equivalence relation $\sim$ as
\begin{equation}
\label{eq: equivachains}
    A \sim B \ \ \Longleftrightarrow \ \ A_{ij}=\lambda _i B_{ij} \lambda _j ^{-1} \ ,
\end{equation}
which is well defined since
\be
\delta_1 A_{ijk} = \lambda_j (\delta_1 B_{ijk}) \lambda_j ^{-1} \ .
\ee
Quotienting by $\sim$ defines $H^1(X,G)$. In physical terms the equivalence relation above is a gauge transformation. $H^1(X,G)$ is not a group, but this does not affect the formalism of \cite{Antinucci:2022vyk} in any way. 

The second issue is that while the twisted sectors of the four-dimensional defects (and the duality defects) are labelled by elements of $G$, the GW operators are labelled by conjugacy classes. Indeed a GW operator for the discrete gauge theory is defined by a singular connection $A$ such that
\be
\label{eq: disorder operator}
\delta_1 A_{ijk} = g \, ,
\ee
around a patch $U_{ijk}$. Since a gauge transformation \eqref{eq: equivachains} by $\lambda \in C^0(X,G)$ on $A$ maps $\delta _1 A_{ijk}$ to $\lambda _j \delta _1 A_{ijk} \lambda _j ^{-1}$  we have to declare that GW operators labelled by elements in the same conjugacy classes are equivalent: thus a GW operator is labelled by a conjugacy class $[g]$ rather than an element of the group.

This fact is reflected on twist defects as follows. For $G$ an abelian group (as in \cite{Antinucci:2022vyk}) the 3d action \eqref{eq: action twist}  defining the twist defect $D[\cT(M)]$ has a 0-form symmetry $G$, which acts as $\cB \rightarrow {{M'}\inv}^\sT \cB$, $\Phi \rightarrow M' \Phi$, $\Gamma\rightarrow M' \Gamma$, $\Psi \rightarrow {{M'}\inv}^\sT \Psi$, where $M'\in G$. Indeed 
\begin{equation}\label{eq: conjugation T}
\cT(M)\rightarrow {M'}^\sT\; \cT(M) \; M'=\cT ({M'}\inv M M')
\end{equation}
which for $G$ abelian is $\cT (M)$. 
For a non-abelian $G$, while the GW operators are labelled by conjugacy classes, the 3d twist defects are labelled by elements $M\in G$. However because of \eqref{eq: conjugation T} the 3d action does not have the $G$ symmetry in the non-abelian case, but the action of $M'\in G$ on $D[\cT (M)]$ produces a different defect:
\bea
\begin{tikzpicture}    \draw (0,0) -- (0,3);
    \draw (0,1) arc (90:-90: 1 and 1);
    \draw (0,2) arc (90:270: 1.5 and 1.5);
    \draw[fill=red] (0,0) circle (0.1);
    \node[below] at (0,0) {$D[\cT]$};
    \node[above] at (0,3) {${M'}\inv M M'$};
    \node[right] at (1,0) {$M'$};
    \node[left] at (-1.5,0) {${M'}\inv$};
    \node[left] at (0,0.5) {$M$};
    \node[right] at (0,1.5) {$M M'$};
    \end{tikzpicture}
\eea
Thus we cannot simply covariantize the action to generate a good operator in the gauged theory. What we have to do, instead, is to sum over all defects which are in the same orbits for the adjoint action of $G$ on itself:\begin{equation}
    D[\cT(M)]\rightarrow D[\cT(M)]/G=\bigoplus _{\widetilde{M}\in [M]} D[\cT (\widetilde{M})]\,.
\end{equation}
We conclude that in the gauged theory also the twist defects are labelled by the conjugacy classes of $G$, and they form a compound with the GW operators defined by the equation \eqref{eq: disorder operator}. A convenient way to represent these operators is to start from the \emph{naked}, non gauge-invariant three-dimensional operators labelled by elements $M\in G$, then acting on it with gauge transformations and summing over them:
\bea    \begin{tikzpicture}    \draw (0,0) -- (0,2);
    \draw[fill=green] (0,0) circle (0.1);
    \node[below] at (0,-0.1) {$GW_{[M]}$};
    \node at (2,0) {$=$};
    \node at (3,-0.15) {$\ds \sum_{M' \in G}$};
    \draw (5,0) circle (1.5);
    \draw (5,0) -- (5,2);
    \draw[fill=black!30!green] (5,0) circle (0.1);
    
    \node[below] at (5,-0.1) {$GW_M$};
    \node[right] at (6.5,0) {$M'$};
    \end{tikzpicture}
\eea
The four-dimensional defects implementing $M,M'$ are precisely the location of the co-dimension one plane orthogonal to the 1-simplices associated with the gauge field $A\in H^1(X_5,G)$, and therefore are transparent in the gauged theory.
Notice that in the bulk, the 3d operators of the gauged theory have two sources of non-invertibility. The first one, which we discussed here and in \cite{Antinucci:2022vyk}, has to do with the appearance of TQFT coefficients and condensates, while the second one comes from the fact that these defects are sum of several defects of the ungauged theory leading to a non-invertibility of orbifold type as in \cite{Bhardwaj:2022lsg, Antinucci:2022eat}.

A gapped boundary of the gauged theory can be described as a non-simple but $G$-invariant boundary in the ungauged theory, tensored with Dirichlet boundary conditions for the $G$ gauge field:
\be
|\rho/G\rangle = \frac{1}{|\text{Stab}(\rho)|}\sum_{M \in G } |\rho_M \rangle \times |A=0\rangle
\ee
Bringing the GW operator onto the gapped boundary liberates the naked GW since four-dimensional surfaces implementing $G$ are absorbed by $|\rho/G\rangle$:
\bea    \begin{tikzpicture}[scale=0.75]     \draw[color = green, line width=2.5] (-2,0) -- (2,0);
      \draw (0,2) circle (1.5);
      \draw (0,2) -- (0,4);
    \draw[fill=black!30!green] (0,2) circle (0.1);
        \node[below] at (2,0) {$\rho/G$};
\node[below] at (0,1.9) {$GW_M$};
\node at (3,1) {\LARGE $ \leadsto $};
    \begin{scope}[shift={(6,0)}]
    \draw (-1.5,0) arc (180:0: 1.5 and 1.5);
    \draw[color = green, line width=2.5] (-2,0) -- (2,0);
    \draw (0,0) -- (0,4);
    \draw[fill= black!30!green] (0,0) circle (0.1);
    \node[below] at (2,0) {$\rho/G$};
    \node[below] at (0,-0.1) {$GW_M$};
    \end{scope}
    \end{tikzpicture} \hspace{6em}
      \begin{tikzpicture}[scale=0.75]
     \draw[color = green, line width=2.5] (-2,0) -- (2,0);
     \draw (0,0) -- (0,2);
    \draw[fill= black!30!green] (0,0) circle (0.1);
    \draw[dashed] (-90:1) -- (30:1) -- (150:1) -- cycle;
    \draw[fill=black] (-90:1) circle (0.05);
    \draw[fill=black] (30:1) circle (0.05);
    \draw[fill=black] (150:1) circle (0.05);
    \node at (-90:1.3) {$3$};
    \node at (30:1.3) {$2$};
    \node at (150:1.3) {$1$};
    \node[above] at (0,2) {$M$};
    \end{tikzpicture} 
    \label{fig: miscGW}
\eea

Alternatively, we can just think of the gauge transformations also being frozen on the boundary, so that $\left(\delta_1 A\right)_{ijk}= M \in G$ is a well defined boundary condition. This is depicted on the right side of \eqref{fig: miscGW}, with $A_{1 3} = A_{2 3} = 1$ and $A_{12}= M$. The last specification should be thought of as a boundary condition for the insertion of the boundary defect. 
Thus we conclude that upon bringing twist defect onto a gapped boundary the non-invertibility of the orbifold type disappears and these operators are labelled by element $M\in G$ instead of conjugacy classes, as we expect from the $4d$ construction.

\section{Applications and Examples}\label{sec: Examples}
We conclude with some explicit computations for $g=2$. The cases of higher genus can be treated similarly, however the number of global variants soon becomes prohibitive and we do not expect any new features to emerge. 
The list of discrete automorphism groups for $g=2$ Riemann surfaces without puntures is the following \cite{breuer2000characters}:
\begin{center}
\begin{tabular}{|c|c|}
\hline
$G(\Omega)$ & $\#$ Moduli   \\
\hline
    $\bZ_{10}$ & 0 \\
    \hline
    $(\bZ_2 \times \bZ_6) \rtimes \bZ_2$ & 0 \\
    \hline
   $ \bZ_{12}\times \bZ_2$ & 0 \\
    \hline
 $   (\bZ_4 \times \bZ_4) \rtimes \bZ_2$ & 0 \\
    \hline
  $  \text{GL}(2,3)$ & 0 \\
    \hline
   $ \bZ_3\times(\bZ_6 \times \bZ_2) \rtimes \bZ_2$ & 0 \\
    \hline
   $ \bZ_2 \times \bZ_4 $ & 1 \\
    \hline
   $ D_8 $ & 1 \\
    \hline
    $\bZ_2 \times    \bZ_6$ & 1 \\
    \hline
   $ D_{12}$ & 1 \\
    \hline
  $  \bZ_2 \times    \bZ_2$ & 2 \\
    \hline
    \end{tabular}
\end{center}
We focus on two representative cases: the largest cyclic group $\bZ_{4g +2} = \bZ_2^C \times \bZ_{2 g +1}$ and the symmetry enhancement from $D_{4g +4}$ to $(\bZ_{2 g + 2} \times \bZ_2) \rtimes \bZ_2$. Both cases are present for every genus. To compute the fusion rules using the discrete gauging description we first determine the matrix $\Phi_\cL^{M\inv}$ which sends $V$ to $M\inv V$ (seen as a member of the quotient $\Sp{(2g, \bZ_N)}/\cP(2g ,\bZ_N)$) via its right action. This is just
\be
\Phi_\cL^{M\inv} = V\inv \; M\inv \; V \, .
\ee
We then decompose it in the standard form as explained in Appendix \ref{app: rep}.\footnote{Notice that the electric boundary is just $\cL = \unit_{2g}$, to topological manipulations there can  be extracted by putting $M\inv$ in standard form.} All the remaining categorical data are extracted by applying the K-formula.

\paragraph{Cyclic group $\bZ_{4g +2}$.} The Riemann surface with such an automorphism group is rather simple to describe, it corresponds to the hyperelliptic curve:
\be
y^2 = x^{2 g +1} - 1 \, .
\ee
The symmetry is comprised by the hyperelliptic involution $\bZ_2^C \; : C y = - y$ and a discrete rotation $\bZ_{2g +1} \; : \rho x = e^{ -\frac{2 \pi i}{2 g + 1} } \; x$. The action on homology cycles is best seen by representing the curve as a branched cover of the complex plane (here for $g=2$):
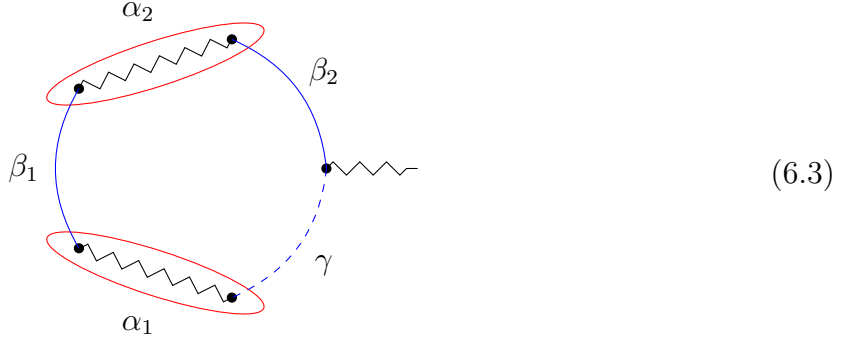
\bea
\begin{tikzpicture}[scale=0.6, rotate=-90]
\coordinate (a1) at (90:3); \coordinate (a2) at (18:3); \coordinate (a3) at (-54:3); \coordinate (a4) at (234:3); \coordinate (a5) at (162:3) ;
\draw[fill=black] (a1) circle (0.1); \draw[fill=black] (a2) circle (0.1); \draw[fill=black] (a3) circle (0.1); \draw[fill=black] (a4) circle (0.1); \draw[fill=black] (a5) circle (0.1);
\draw[decorate, decoration= zigzag] (a2) -- (a3); \draw[decorate, decoration= zigzag] (a4) -- (a5); \draw[decorate, decoration= zigzag] (a1) -- (0,5);
\draw[rotate around={-18: ($0.5*(a2) + 0.5*(a3)$)}, color=red] ($0.5*(a2) + 0.5*(a3)$) ellipse (0.5 and 2.5);
\draw[rotate around={18: ($0.5*(a4) + 0.5*(a5)$)}, color=red] ($0.5*(a4) + 0.5*(a5)$) ellipse (0.5 and 2.5);
\draw[color=blue] (a1) to[bend right=30] (a5);
\draw[color=blue, dashed] (a1) to[bend left=30] (a2);
\draw[color=blue] (a3) to[bend left=30] (a4);
\node at ($0.75*(a2) + 0.75*(a3)$) {$\alpha_1$};
\node at ($0.75*(a4) + 0.75*(a3)$) {$\beta_1$};
\node at ($0.75*(a5) + 0.75*(a4)$) {$\alpha_2$};
\node at ($0.75*(a1) + 0.75*(a5)$) {$\beta_2$};
\node at ($0.75*(a2) + 0.75*(a1)$) {$\gamma$};
\end{tikzpicture}
\eea
The cycles are related to the standard basis as follows: $\alpha_i = A_i$, $\beta_1 = B_1 - B_2$, $\beta_2= B_2$. Charge conjugation $C$ interchanges the two sheets, reversing the orientation of cycles, while $\rho$ corresponds to a discrete clockwise rotation. They correspond to matrices $M \in \Sp{(4, \bZ)}$
\be
M_C = \mat{ -1 & 0 & 0 & 0 \\ 0 & -1 & 0 & 0 \\ 0 & 0 & -1 & 0 \\ 0 & 0 & 0 & -1 } \, , \ \ M_\rho = \mat{0 & 0 & 1 & -1 \\ 0 & 0 & 0 & 1 \\ -1 & 0 & 1 & 0 \\ -1 & -1 & 1 & 0 }\,.
\ee
In this example the only interesting defect is $M_\rho$, since $C$ leaves all the Lagrangian lattices $\cL$ invariant and is thus invertible. It can be checked that, for all $\cL$:
\be
\fD_\cL^{M_\rho} \times {\fD_\cL^{M_\rho}}^\dagger = \cC^{\cA} \, ,
\ee
where $\cA$ is subgroup of the one-form symmetry being gauged by $\Phi_\cL^{M_\rho\inv}$. For $g=2$ this can either be the full $\bZ_N^2$, a one dimensional subgroup $\cA$ or nothing at all, in which case the defect is invertible.

The fusion rules for $M_\rho$ with itself are more interesting. For $g=2$ there can be several patterns, depending on $r(M_\rho)$ and $r(M_\rho^2)$. The allowed patterns are given in the Table below:
\begin{center}
$
\begin{array}{| c | c | c |}
\hline
   r(M_\rho)  & r(M_\rho^2) & \fD_\cL^{M_\rho} \times \fD_\cL^{M_\rho}  \\[0.25em] \hline
     &  &   \\[-1.25em]
     2 & 2 & \cN^{2,\pm} \; \fD_\cL^{M_\rho^2} \\ \hline
     2 & 1 & \cC^{\bZ_N} \;  \cN^{1, \pm} \;   \fD_\cL^{M_\rho^2} \\ \hline
     2 & 0 & \cC^{\bZ_N^2} \; \cU_\cL^{M_\rho^2} \\ \hline
     1 & 2 & \fD_\cL^{M_\rho^2} \\ \hline
     1 & 1 &  \cN^{1,\pm} \; \fD_\cL^{M_\rho^2} \\ \hline
     1 & 0 & \cC^{\bZ_N} \; \cU_\cL^{M_\rho^2} \\ \hline
     0 & 0 &  \cU_\cL^{M_\rho^2}\\  \hline
\end{array}
$
\end{center}
As explained in appendix \ref{app: quadratic} we only have two choices $\cN^{r, \; \pm}$ for the TQFT coefficients at a given rank. In our conventions $\cN^{r,+}$ is represented by $\left(\cA^{N,1}\right)^{r}$ while $\cN^{r,-}$ by $\left(\cA^{N,1}\right)^{r-1} \times \cA^{N, q'}$, $q'$ not being a perfect square.

Notice that, even if all three defects are non-invertible, they can fuse as an invertible symmetry sometimes. We also give in appendix \ref{app: Tables} the full multiplication table for the case $g=2$, $N=3$, which we have checked from both the 5d Symmetry TFT and the 4d QFT perspective. Notice that the agreement of the results of the two completely independent computations is a highly non-trivial check of the proposal of \cite{Antinucci:2022vyk} for the holographic dual of self-duality symmetries.

\paragraph{Dihedral group $D_{4g+4}$ and symmetry enhancement.} The second example is the simplest non-abelian group $D_{4g+4}$. This example also enjoys two other features of non-invertible symmetries in class $\cS$ theories, namely the presence of moduli spaces on which the symmetry is realized and its enhancement. 
As before we can represent the surface hosting such automorphism group by an hyperelliptic curve\footnote{This is true for even genus, if $g$ is odd instead
\be
y^2 = x \left( x^g -\lambda \right)\left( x^g - 1/\lambda \right)
\ee
and the dihedral group is $D_{4g}$.
}
\be
y^2 = \left( x^{g +1} - \lambda \right)\left( x^{g+1} - 1/\lambda \right) \, . 
\ee
The symmetry is generated by the hyperelliptic involution $\bZ_2^C$, a rotation $\bZ_{g+1}^{t}\; : t x = e^{-\frac{2 \pi i}{g +1}}x$ and a reflection $\bZ_2^r \; : r(y, \;x) = (y \; x^{-(g + 1)}, \; x^{-1})$. It is simple to show that they combine into a dihedral group $D_{4g + 4}$. The symmetry is enhanced at the special point $\lambda=i$ by a further $\bZ_2$ symmetry $\bZ_2^\sigma \; : \sigma x = - x$ to the group $(\bZ_{2g +2} \times \bZ_2^C) \rtimes \bZ_2^r$. $\sigma$ should be thought as an emergent $S$-duality.
Again the action of the symmetry group on homology is easier to visualize by employing a branched cover
\bea
\begin{tikzpicture}[scale=0.8]
    \coordinate (a1) at (0:2); \coordinate (b1) at (0:3); \coordinate (a2) at (-120:2); \coordinate (b2) at (-120:3); \coordinate (a3) at (-240:2); \coordinate (b3) at (-240:3); \coordinate (o) at (0,0);
    \draw[fill=black] (a1) circle (0.1);     \draw[fill=black] (a2) circle (0.1);
     \draw[fill=black] (a3) circle (0.1);
     \draw[fill=black] (b1) circle (0.1);
     \draw[fill=black] (b2) circle (0.1);
     \draw[fill=black] (b3) circle (0.1);
     \draw[decorate, decoration= zigzag] (a1)--(b1);      \draw[decorate, decoration= zigzag] (a2)--(b2);
      \draw[decorate, decoration= zigzag] (a3)--(b3);
            \draw[color=red,rotate around={-120:($0.5*(a2) + 0.5*(b2)$)}] ($0.5*(a2) + 0.5*(b2)$) ellipse (0.8 and 0.3);
      \draw[color=red,rotate around={120:($0.5*(a3) + 0.5*(b3)$)}] ($0.5*(a3) + 0.5*(b3)$) ellipse (0.8 and 0.3);
       \draw[color=red, dashed, rotate around={0:($0.5*(a1) + 0.5*(b1)$)}] ($0.5*(a1) + 0.5*(b1)$) ellipse (0.8 and 0.3);
\draw[color = blue] (a3) to[bend right = 30] (b2);
\draw[color = blue, dashed] (a2) to[bend right = 30] (b1);
\draw[color = blue] (a1) to[bend right = 30] (b3);

\node at ($1.2*(b2)$) {$\alpha_1$};
\node at ($1.2*(b3)$) {$\alpha_2$};
\node at ($1.2*(b1)$) {$\alpha_{0}$};
\node at ($1.2*(a2)+ 1.2*(a3)$) {$\beta_1$};
\node at ($1.2*(a3)+ 1.2*(a1)$) {$\beta_2$};
\node at ($1.2*(a2)+ 1.2*(a1)$) {$\beta_0$};

\begin{scope}[shift={(11,0)}]
         \def\th{30};
    \def\al{-60};
    \def\rr{2.5};
    \coordinate (a1) at ($(\th:\rr)$);  \coordinate (a2) at ($(\th + \al:\rr)$);  \coordinate (a3) at ($(\th + 2*\al:\rr)$);  \coordinate (a4) at ($(\th + 3*\al:\rr)$);  \coordinate (a5) at ($(\th + 4*\al:\rr)$);  \coordinate (a6) at ($(\th + 5*\al:\rr)$);
    \coordinate (b1) at ($(-\th:\rr)$);  \coordinate (b2) at ($(-\th + \al:\rr)$);  \coordinate (b3) at ($(-\th + 2*\al:\rr)$);  \coordinate (b4) at ($(-\th + 3*\al:\rr)$);  \coordinate (b5) at ($(-\th + 4*\al:\rr)$);
    \draw[fill=black] (a1) circle (0.1);
        \draw[fill=black] (a2) circle (0.1);
    \draw[fill=black] (a3) circle (0.1);
        \draw[fill=black] (a4) circle (0.1);
    \draw[fill=black] (a5) circle (0.1);
    \draw[fill=black] (a6) circle (0.1);
\draw[decorate, decoration= zigzag] (a1) -- (a2);
\draw[decorate, decoration= zigzag] (a3) -- (a4);
\draw[decorate, decoration= zigzag] (a5) -- (a6);

\draw[color=red, dashed, rotate around={90:($0.5*(a1) + 0.5*(a2)$)}] ($0.5*(a1) + 0.5*(a2)$) ellipse (2 and 0.3);
\draw[color=red, rotate around={-30:($0.5*(a3) + 0.5*(a4)$)}] ($0.5*(a3) + 0.5*(a4)$) ellipse (2 and 0.3);
\draw[color=red, rotate around={-150:($0.5*(a5) + 0.5*(a6)$)}] ($0.5*(a5) + 0.5*(a6)$) ellipse (2 and 0.3);

\draw[color=blue, dashed] (a2) to[bend left =30] (a3);
\draw[color=blue] (a4) to[bend left =30] (a5);
\draw[color=blue] (a6) to[bend left =30] (a1);
\node at ($0.65*(a1) + 0.65*(a2)$) {$\alpha_0$};
\node at ($0.65*(a3) + 0.65*(a4)$) {$\alpha_1$};
\node at ($0.65*(a5) + 0.65*(a6)$) {$\alpha_2$};

\node at ($0.65*(a2) + 0.65*(a3)$) {$\beta_0$};
\node at ($0.65*(a4) + 0.65*(a5)$) {$\beta_1$};
\node at ($0.65*(a6) + 0.65*(a1)$) {$\beta_2$};

\end{scope}

\end{tikzpicture}
\eea
The left picture is the curve for $\lambda \in \bR^+$, while the right one represents the special point $\lambda = i$. The actions of $t$ and $\sigma$ are just rotations, while $r$ flips the picture around the real axis while also interchanging the two ends of each cut. For $g=2$ their matrix representation is
\be
M_t = \mat{0 & 1 & 0 & 0 \\ -1 & -1 & 0 & 0 \\ 0 & 0 & -1 & 1 \\ 0 & 0 & -1 & 0} \, , \ \ M_r = \mat{ 0 & 1 & 0 & 0 \\ 1 & 0 & 0 & 0 \\ 0 & 0 & 0 & 1 \\ 0 & 0 & 1 & 0 } \, , \ \  M_\sigma = \mat{ 0 & 0 & 1  & -1 \\ 0 & 0 & 0 & 1 \\ -1 & 0 & 0 & 0 \\ -1 & -1 & 0 & 0 }
\ee
In this case the duality group is non-abelian, and one might wonder whether the categorical structure of its fusion rules is also non-commutative. The answer is positive. For example consider $N=3$ and 
\be \cL =\mat{ 0 & 0 \\ 0 & 1 \\ 1 & 0 \\ 1 & 1} \ \ \ \ \ \ \fD_\cL^{M_r} \; \times \; \fD_\cL^{M_\sigma} = \cN^{2, + }  \; \fD_\cL^{M_{r \; \sigma}} \ \ , \ \ \fD_\cL^{M_\sigma} \; \times \; \fD_\cL^{M_r} = \cN^{2 , -} \; \fD_\cL^{M_{\sigma \; r}} \, . \ee 
The non-commutativity may also involve condensation defects, for example 
\be
\cL=\mat{0 & 1 \\ 1 & 1 \\ 1 & 0 \\ 0 & 1} \ \ \ \ \ \ \fD_\cL^{M_r} \; \times \; \fD_\cL^{M_\sigma} = \cN^{2, - }  \; \fD_\cL^{M_{r \; \sigma}} \ \ , \ \ \fD_\cL^{M_\sigma} \; \times \; \fD_\cL^{M_r} = \cC^{\bZ_N^2} \; \cU_\cL^{M_{\sigma \; r}} \, . 
\ee
While this happens, associativity must still hold. Let us again consider the second example above and compute $\fD_\cL^{M_r} \times \fD_\cL^{M_\sigma} \times \fD_\cL^{M_r} = \fD_\cL^{M_{ r \; \sigma \; r}}$, in the two fusion channels we get
\be
\fD_\cL^{M_r} \times \fD_\cL^{M_\sigma} \times \fD_\cL^{M_r} = \begin{cases}
    &\left( \cN^{2, -}\right)^2 \; \fD_\cL^{M_{r \; \sigma \; r}} \\
    & \cC^{\bZ_N^2} \; \fD_\cL^{M_{r \; \sigma \; r}} 
\end{cases}
\ee
To compare the expressions notice that, for $N=3$ $(\cN^{2, -})^2 = \cN^{4,+} = (\bZ_N)^2$ while the condensation can be absorbed by the $\fD_\cL^{M_{r \; \sigma \; r}}$ defect as it is full rank, leaving behind a $(\bZ_N)^2$ partition function.
We give a table of the non-commuting cases in Appendix \ref{app: Tables}.

\section{Conclusions}\label{sec: conclusions}
The main goal of this paper was understanding non-invertible duality defects in theories with an extended 1-form symmetry $(\bZ_N)^g$.
These are naturally realized in 4d SCFT of class $\cS$ whose Riemann surface $\Sigma_g$ has a nontrivial automorphism group $G(\Omega)$ which acts as self-dualities.
Due to the large 1-form symmetry group the structure of the non-invertible defects is more intricate than \eg \ the case of $\cN=4$ SYM. Given two elements $M_1$ and $M_2$ in $G(\Omega)$ and a choice $\cL$ of global variant the generic fusion between duality defects takes the form
\be
 \fD_\cL^{M_1} \times \fD_\cL^{M_2} = \cN ^{r, \pm} \; \cC^\cA  \; \fD_\cL^{M_{1,2}} \, ,
\ee
where $\cN^{r , \pm}$ are decoupled TQFTs and $\cC^{\cA}$ condensation defects for $\cA \subset (\bZ_N)^g$.
We have given two ways to understand this structure: either from a purely QFT perspective or from $5$d TQFT description. In the former it can be understood by considering the algebra of the group $\Sp{(2g, \bZ_N)}_T$ of discrete topological manipulations $\Phi$ on half space, while in the latter it descends from the description of twist defects in a certain 5d Chern-Simons theory, which become liberated as special points of the gravitational moduli space where $G(\Omega)$ remains un-Higgsed. This generalizes the previous analysis of \cite{Antinucci:2022vyk}.
The validity of our approach is shown in various concrete examples for $N=3$ and $g=2$, pointing out some interesting new properties such as the non-commutativity of the non-invertible symmetry algebra.

Finally, we have given slick way to derive the fusion rules by analyzing the action of the non-invertible defects on genuine line operators. This has led us to introduce the concept of ``rank'' of a non-invertible symmetry. 

Let us close by commenting on open questions and natural generalizations of our results.
Some natural extensions are the addition of punctures on the Riemann surface --- both regular and irregular --- and the study of class $\cS$ theories of $D$ and $E$-type. The one-form symmetry structure in these cases has been studied \eg \ in \cite{Bhardwaj:2021pfz,Bhardwaj:2021mzl}. We expect that, as long as only regular punctures are involved, our methods should extend without great novelties. A second question concerns 't Hooft anomalies and the possibility of gauging these duality symmetries. This should have an interpretation in both the 4d SCFT and the 5d Symmetry TFT. Finally, we are left wandering whether preserving these symmetries also gives rise to new exotic RG flows. It is simple to prove, at least under some mild assumptions\footnote{The proof mimics the arguments of \cite{Choi:2021kmx}.}, that these symmetries cannot generically be realized by an SPT in the IR. It would be interesting to study the existence of such RG flows further.\\[2em]
\paragraph{Acknowledgements.} We thank V. Bashmakov, M. Del Zotto, J. Kaidi, K. Ohmori, S. Schafer-Nameki and Y. Wang for useful discussions. We especially thank F. Benini for discussions and collaboration on a related project.
We gratefully acknowledge support from
the Simons Center for Geometry and Physics, Stony Brook University, where some of the
research for this paper was performed. The authors are partially supported by the INFN ``Iniziativa Specifica ST\&FI''. A.A., C.C. and G.R.
are supported by the ERC-COG grant NP-QFT No. 864583 ``Non-perturbative dynamics
of quantum fields: from new deconfined phases of matter to quantum black holes'', by the
MIUR-SIR grant RBSI1471GJ, and by the MIUR-PRIN contract 2015 MP2CX4.

\appendix

\section{Matrix representation of $\Sp{(2g, \, \bZ_N)}_T$}\label{app: rep}
In this appendix we describe the matrix representation of generic topological manipulations $\Phi \in \Sp{(2g, \; \bZ_N)}$. The matrices generating the parabolic group are
\be
\tau(S) = \mat{ \unit & S \\ 0 & \unit} \, , \ \ \ \nu(U) = \mat{ {U\inv}^\sT & 0 \\ 0 & U } \, .
\ee
Notice that, under right composition\footnote{To follow the notation in the main paper, topological manipulations act multiplicatively from the right on the global variant parameterized by a matrix $V\in \text{Sp}(2g,\bZ_N)$.}
\be
\tau(S) \; \nu(U) = \mat{  {U\inv}^\sT & {U\inv}^\sT \; S \\ 0 & U } = \nu(U) \; \tau( {U\inv}^\sT \; S \; U\inv   )
\ee
as it should be. 
A generic gauging matrix $\sigma(C_\cA)$, satisfying 
\be
\sigma(C_\cA) \; \sigma(C_\cA) = \nu( \textbf{C}_\cA ) \, ,  \ \ \ \textbf{C}_\cA = -C_\cA \; C_\cA^* + C_{\Tilde{\cA}} \; C_{\Tilde{\cA}}^*
\ee
can be constructed in the following way: start with the rank $r$ projector $P = \sum_{j=1}^r E_j$ and its complementary $P^\perp = \unit - P$. Define the matrix:
\be
\sigma(P) = \mat{ P^\perp & - P \\ P & P^\perp }
\ee
which describes the gauging of $E_1 \smallvee E_2 \smallvee ... \smallvee E_r$. It is an $\Sp{(2g, \bZ_N)}_T$ matrix and $\sigma(P)^2 = \mat{ P^\perp - P & 0 \\ 0 & P^\perp - P }$. Consider 
\be
\sigma(C_\cA) \equiv \nu(u\inv) \; \sigma(P) \; \nu(u)  =\mat{ {u\inv}^\sT \; P^\perp \; u^\sT & - {u\inv}^\sT \; P \; u\inv  \\ u \; P \; u^\sT & u \; P^\perp \; u\inv  } \, ,
\ee
where $u = \left( C_\cA \; \vert \; C_{\Tilde{\cA}} \right)$. Then
$u P = \left( C_\cA \; \vert \; 0 \right)$, $u P^\perp = \left( 0 \; \vert \; C_{\Tilde{\cA}} \right)$, while
\be
P u\inv = \mat{ C_{\cA}^*  \\ 0  }  \quad P^\perp u\inv = \mat{0 \\  C_{\Tilde{\cA}}^*}\,.
\ee
Then
\be
    \sigma(C_\cA)^2 = \mat{ {u\inv}^\sT \; ( P^\perp - P ) \; u^\sT & 0 \\ 0 & u (P^\perp - P) u\inv  } = \mat{ {\textbf{C}_\cA\inv}^\sT & 0 \\ 0 & \textbf{C}_{\cA}  } \, .
\ee
We can also write down the matrix corresponding to $\sigma(C_\cA) \; \tau({S_\cA}^\cA) \equiv \Phi_{\cA \, , S_\cA}$:\footnote{We think of $S_\cA$ as a matrix with non-zero entries only in the upper-left $r \times r$ corner.}
\be
\Phi_{\cA \, , S_\cA} = \mat{ {u\inv}^\sT (P^\perp +S_\cA ) u^\sT & - {u\inv}^\sT \; P \; u\inv  \\ u \; P \; u^\sT & u \; P^\perp \; u\inv}\,.
\ee
From these definitions it is possible to reconstruct the full algebra of $\Sp(2g\, , \bZ_N)_T$ barring the central extension. We also want to prove the standard-form decomposition for elements $\Phi \in \Sp{(2g, \bZ_N)}_T$. That is, we want to write \footnote{Here matrices are written on the left, but composition should be understood on the right.}:
\be
\Phi= \nu(E) \; \tau(S') \; \sigma(C_\cA) \; \tau(S) \, , \ \ S = {S_\cA}^\cA = (u\inv)^\sT \; P \; s \; P u\inv \, .
\ee
 We parametrize $\Phi= \mat{ A & B \\ C & D}$. A short computation shows that $C = u \; P \; u^\sT \; {E\inv}^\sT$, so the matrices $u$ and $E$ can be extracted by computing the Smith Normal Form of $C$. Having done this we find a matrix $Y_\cA = \sigma(C_\cA) \; \tau(S)$ such that $\Phi \; Y_\cA\inv$ is parabolic. This can be done if the equation\footnote{This comes from setting the bottom left block of $\Phi \; Y_\cA\inv$ to zero.}
\be
P \; s \; P u^\sT - P \; u^\sT \; A \; E^\sT = 0 \mod N \; 
\ee
has solutions. Imposing that $\Phi$ is symplectic we get that $A E^\sT = Q \; u \; P \; u^\sT$ for some matrix $Q$. One finally sets $s = u^\sT \; Q \; u$ to solve the equation. With this procedure it is possible to put any discrete manipulation $\Phi$ into the standard form.

\section{Composition laws for topological manipulations}\label{app: composition comp}
\paragraph{K-formula for $g=1$.} Let us first derive the K-formula for the $g=1$ case:
\be
\left[\sigma \; \tau(k) \; \sigma Z \right](B) = \sum_{b , \, c \; \in \; H^2(X, \,\bZ_N)} \exp\left( \frac{2 \pi i}{N} \int b \cup (c + B) + \frac{k}{2} \fP(b) \right) \; Z(c) \, .
\ee
We change variables $b \to b - k\inv (c + B)$. This cancels the minimal coupling between $b$ and $c$ fields, leaving us with
\bea
&= Y_k \; \exp\left(-\frac{2 \pi i k\inv}{2 N} \fP(B)\right) \; \sum_{c \in H^2(X , \bZ_N)} \exp\left( \frac{2 \pi i}{N} \int c \cup (- k\inv B) - \frac{k\inv}{2} \fP(c) \right) \; Z(c) \\
&= Y_k \left[ \nu(-k\inv) \; \tau(-k) \; \sigma \; \tau(-k\inv) Z  \right] (B) \,
\eea
where the central extension is
\be
Y_k = \sum_{b \in H^2(X, \, \bZ_N)} \exp\left( \frac{2 \pi i k}{2 N} \fP(b)  \right)\,.
\ee
\paragraph{K-formula for generic $g$.}
Let us now treat the general case of $\sigma(C_\cA) \; \tau(S) \; \sigma(C_\cB)$. First, we want to restrict to the case in which $S=(S_\cA)^\cA$. This can be done straightforwardly by expanding the quadratic form $\fP^S$ leading to the identity
\be
\sigma(C_\cA) \tau(S) = \tau( {S_{\Tilde{\cA}}}^{\Tilde{\cA}} ) \; \nu(V_\cA) \; \sigma(C_\cA) \; \tau( {S_\cA}^\cA  ) \, , \ \ \ V_\cA = C_\cA \; C_\cA^* + C_{\Tilde{\cA}} \; C_{\Tilde{\cA}}^* + C_\cA \; S_{\cA \Tilde{\cA}} \; C_{\Tilde{\cA}}^* \, .
\ee
Here $\Tilde{\cA}$ denotes the complement of $\cA$ and $C_{\Tilde{\cA}}$ the corresponding matrix.

We then assume safely that $S= {S_\cA}^\cA$. Secondly, we want to change our basis of generators so that $C_\cA = \left( C_{\cA'} \; \vert \; C_\cC \right)$, $C_\cB= \left( C_{\cB'} \; \vert \; C_\cC \right)$ with $\cC = \cA \smallwedge \cB$. This can be implemented through right multiplication by a $\text{GL}(r(\cA) , \bZ_N)$ matrix $u_\cA$ and similarly for $\cB$. Using the definition of $\sigma$ it is simple to see that the two are related by $\nu$ transformations. With this in mind we can assume that all the matrices are already given in their ``split'' form.

The double gauging reads, explicitly
\bea \label{eq: gaugingwithtorsiong}
&\left[\sigma(C_\cA) \; \tau({S_\cA}^\cA) \; \sigma(C_\cB) Z \right](B) = \\
&=\sum_{\substack{\alpha_{\cA'} \, , \alpha(\cC) \\ \\ \beta_{\cB'} \, , \beta_\cC }} \exp \frac{2 \pi i}{N} \int \left( \alpha_{\cA'} \cup B_{\cA'} + \beta_{\cB'} \cup B_{\cB'} + \alpha_\cC \cup ( \beta_\cC + B_\cC) + \frac{1}{2} \fP^{S}(C_{\cA'} \alpha_\cA + C_{\cC} \alpha_\cC) \right)   \\
 & \ \ \ \ \ \ \ \ \ \ \ \ \ \ \ \ \ \ \ \ \ \ \ \ \ \ \ \ \ \ \ \ \ \ \ \ \ \ \ \ \ \ \ \ \ \ \ \ \ \ \ \ \ \  \times Z\left( C_{\cA'} \alpha_{\cA'} + C_{\cB'} \beta_{\cB'} + C_\cC \beta_\cC + \Tilde{C} \Tilde{B} \right)\,, 
\eea
where we denoted $\Tilde{C}$ the matrix associated to the complement of the join of $\cA$ and $\cB$.

The quadratic function $\fP^{S}$ expands as $\fP^{S_\cC}(\alpha_\cC) + \fP^{S_{\cA'}}(\alpha_{\cA'}) + 2 \alpha_\cC \cup S_{\cC \; \cA'} \alpha_{\cA'}$.
If $S_\cC$ is invertible we redefine:
\be
\alpha_\cC \to \alpha_\cC - S_\cC\inv ( \beta_\cC + B_\cC + S_{\cC \; \cA'} \; \alpha_{\cA'}) \, ,
\ee
eliminating the linear couplings for $\alpha_\cC$. Expanding:
\bea
\eqref{eq: gaugingwithtorsiong} &= \left( \sum_{\alpha_{\cC}} \exp\left( \frac{2 \pi i}{2 N} \fP^{S_\cC} (\alpha_\cC) \right) \right) \; \exp\left( - \frac{2 \pi i}{2 N} \int \fP^{{S_\cC}\inv}(B_\cC) \right) \times \\
& \times \sum_{\substack{\alpha_{\cA'} \\ \beta_{\cB'} \, , \beta_\cC }} \exp\left( \frac{2 \pi i}{N} \int \alpha_{\cA'} \cup (B_{\cA'} - S_{\cA' \; \cC} \; S_\cC^{\inv} B_\cC) - \beta_\cC \cup {S_\cC}\inv \; B_\cC + \beta_{\cB'} \cup B_{\cB'} \right) \times \\
& \times \exp\left(\frac{2 \pi i}{2 N} \int \fP^{S_{\cA'} - S_{\cA' \; \cC} \; S_\cC\inv \; S_{\cC \; \cA'}}(\alpha_{\cA'}) - \fP^{{S_\cC}\inv}(\beta_\cC) - \alpha_{\cA'} \cup S_{\cA' \; \cC} \; {S_{\cC}}\inv \; \beta_\cC \right) \times \\
& \times Z\left( C_{\cA'} \; \alpha_{\cA'} + C_{\cB'} \; \beta_{\cB'} + C_{\cC} \; \beta_\cC + \Tilde{C} \; \Tilde{B} \right)\,.
\eea
The new torsion is thus given by the matrix
\be
- X_{\cA' \; \cB' \; \cC }= {S_{\cA'}}^{\cA'} - \left(S_{\cA' \; \cC} \; S_\cC\inv \; S_{\cC \; \cA'}\right)^{\cA'} - \left(S_{\cA' \; \cC} \; {S_{\cC}}\inv \right)^{\cA' \; \cC} -  \left( {S_{\cC}}\inv S_{\cC \; \cA'} \right)^{\cC \; \cA'} - {{S_\cC}\inv}^\cC
\ee
while to get the correct couplings we must perform a redefinition on $B= C_{\cA'} \; B_{\cA'} + C_{\cB'} \; B_{\cB'} + C_\cC \; B_\cC + \Tilde{\cC} \; \Tilde{B} $ by a matrix:
\be
U_{\cA' \; \cB' \; \cC} =  C_{\cA'} \left( C_{\cA'}^* - S_{\cA' \; \cC} \; S_\cC\inv C_{\cC}^*  \right) - C_\cC \; S_\cC\inv \; C_\cC^* + C_{\cB'} \; C_{\cB'}^* + \Tilde{C} \; \Tilde{C}^* \, .
\ee
This shows that:
\be
\sigma(C_\cA) \; \tau((S_\cA)^\cA) \; \sigma(C_\cB) = Y_{S_\cC} \; \tau(- {S_{\cC}\inv}^\cC) \; \nu( U_{\cA' \; \cB' \; \cC}) \; \sigma( C_{\cA'} \; \vert \; C_{\cB'} \; \vert \; C_\cC  ) \; \tau(- X_{\cA' \; \cB' \; \cC}) \, .
\ee

Now we consider the case in which $S_\cC$ has a kernel. Since we work at $g=2$ we only consider the case in which (I) $S_\cC=0$ or (II) $C_\cA = C_\cB = \unit$ and $S_\cC$ has a kernel $\cK$.
\paragraph{(I)} Since $S_\cC=0$ $\alpha_\cC$ enters the equation linearly and imposes a constraint:
\be
\beta_\cC = - (B_\cC + S_{\cC \cA'} \; \alpha_{\cA'}) \, .
\ee
Letting $C_\cD = (C_{\cA'} - C_\cC \; S_{\cC \cA'}) $ we find:
\bea
\eqref{eq: gaugingwithtorsiong} &= \sum_{\alpha_{\cA'} \, , \beta_{\cB'}} \exp\left( \frac{2 \pi i}{N} \int \alpha_{\cA'} \cup B_{\cA'} + \beta_{\cB'} \cup B_{\cB'} +\frac{1}{2} \fP^{S_{\cA'}}(\alpha_{\cA'}) \right) \times \\
& \times Z\left( C_{\cD} \; \alpha_{\cA'} + C_{\cB'} \; \beta_{\cB'} - C_{\cC} \; B_{\cC} + \Tilde{C} \; \Tilde{B}  \right)\,.
\eea
This is the partition function of a theory in which $C_{\cD} \; \vert \; C_{\cB'}$ is gauged with torsion ${S_{\cA'}}^\cD$. The background is:
\be
B' = C_\cD \; B_{\cA'} + C_{\cB'} \; B_{\cB'} - C_{\cC} \; B_\cC + \Tilde{C} \; \Tilde{B} \, 
\ee
and is obtained for the original one by a transformation
\be
U_{\cA \; \cB} = C_{\cD} \; C_{\cA'}^* + C_{\cB'} \; C_{\cB'}^* - C_{\cC} \; C_{\cC}^* + \Tilde{\cC} \; \Tilde{\cC}^* \, .
\ee
Therefore in case \textbf{(I)} we have
\be
\sigma(C_\cA) \; \tau((S_\cA)^\cA) \; \sigma(C_\cB) = \nu(U_{\cA \; \cB}) \; \sigma( C_{\cD} \; \vert \; C_{\cB'} ) \; \tau( (S_{\cA'})^{\cD} ) \, .
\ee
\paragraph{(II)} Let $\cK$ be the kernel of $S$. The decompose the dynamical field $\beta = C_{\cK} \beta_\cK + C_{\Tilde{\cK}} \; \beta_{\Tilde{\cK}}$. The discrete gauging is:
\bea
\left[\sigma(\unit) \; \tau(S) \; \sigma(\unit) \; Z\right](B) &= \sum_{\substack{\alpha_\cK \, , \alpha_{\Tilde{\cK}} \\ \beta_\cK \, , \beta_{\Tilde{\cK}} }} \; \exp\left( \frac{2 \pi i}{N} \int \alpha_{\cK} \cup (\beta_{cK} + B_\cK) + \alpha_{\Tilde{\cK}} \cup (\beta_{\Tilde{\cK}} + B_{\Tilde{\cK}}) + \frac{1}{2} \fP^{S_{\Tilde{\cK}}}(\alpha_{\Tilde{\cK}}) \right) \times \\
&\times Z\left( C_\cK \; \beta_\cK + C_{\Tilde{\cK} } \; \beta_{\Tilde{\cK}} \right) \, .
\eea
Since $\alpha_\cK$ appears linearly we integrate it out, fixing $\beta_\cK = - B_\cK$. Thus finding
\be
\sigma(\unit) \; \tau(S) \; \sigma(\unit) = \nu\left(- C_\cK \; C_\cK^* + C_{\Tilde{\cK}} \; C_{\Tilde{\cK}}^* \right) \; \sigma(C_{\Tilde{\cK}}) \; \tau({S_{\Tilde{\cK}}}^{\Tilde{\cK}}) \; \sigma(C_{\Tilde{\cK}})\,.
\ee
Applying the K-formula the second term becomes
\be
\sigma(C_{\Tilde{\cK}}) \; \tau({S_{\Tilde{\cK}}}^{\Tilde{\cK}}) \; \sigma(C_{\Tilde{\cK}}) = Y_{S_{\Tilde{\cK}}} \; \tau\left( - {S_{\Tilde{\cK}}\inv}^{\Tilde{\cK}} \right) \; \nu\left( C_\cK \; C_\cK^* - C_{\Tilde{\cK}} \; S_{\Tilde{\cK}}\inv \; C_{\Tilde{\cK}}^* \right) \; \sigma(C_{\Tilde{\cK}}) \; \tau\left( - {S_{\Tilde{\cK}}\inv}^{\Tilde{\cK}} \right) \,.
\ee
Defining $U=- C_\cK \; C_\cK^* + C_{\Tilde{\cK}} \; C_{\Tilde{\cK}}^*$ and $V= C_\cK \; C_\cK^* - C_{\Tilde{\cK}} \; S_{\Tilde{\cK}}\inv \; C_{\Tilde{\cK}}^*$ we have $V^{-1} = C_\cK \; C_\cK^* - C_{\Tilde{\cK}} \; S_{\Tilde{\cK}} \; C_{\Tilde{\cK}}^*$, $V\; U = W= - \left( C_\cK \; C_\cK^* + C_{\Tilde{\cK}} \; S_{\Tilde{\cK}}\inv \; C_{\Tilde{\cK}}^* \right) $ and $ (V\inv)^\sT \; {S_{\Tilde{\cK}}\inv}^{\Tilde{\cK}} \; V\inv = {S_{\Tilde{\cK}}}^{\Tilde{\cK}}  $. The final formula is
\be
\sigma(\unit) \; \tau(S) \; \sigma(\unit) = Y_{S_{\Tilde{\cK}}} \; \nu(W) \; \tau(- {S_{\Tilde{\cK}}}^{\Tilde{\cK}}) \; \sigma(C_{\Tilde{\cK}}) \; \tau(-  {S_{\Tilde{\cK}}\inv}^{\Tilde{\cK}} ) \, .
\ee

\paragraph{Composing $\sigma(C_\cA)$ and $\nu(U)$} Lastly we need to understand the composition
\be \label{eq: sigmanu}
\left[\sigma(C_\cA) \; \nu(U) \; Z \right](B) = \sum_{\alpha_\cA} \exp\left( \frac{2 \pi i}{N} \int \alpha_\cA \cup B_\cA \right) \; Z\left( U\left(C_\cA \; \alpha_\cA + C_{\Tilde{\cA}} \; B_{\Tilde{\cA}}\right) \right) \,.
\ee
Defining $C_{\cA_U} = U C_\cA$ and $C_{\Tilde{\cA}_U} = U C_{\Tilde{\cA}}$\footnote{The duals instead transform with the inverse matrix, e.g. $C_{\cA_U}^* = C_{\cA} U \inv$.} we can write this as:
\be
\eqref{eq: sigmanu} = \left[ \sigma(C_{\cA_U}) \; Z \right] (B'= C_{\cA_U} \; B_\cA + C_{\Tilde{\cA}_U} \; B_{\Tilde{\cA}})
\ee
The original couplings were instead $B= C_\cA \; B_\cA + C_{\Tilde{\cA}} \; B_{\Tilde{\cA}}$, so we need to compose with $\nu(U)$ 
\be
\sigma(C_\cA) \; \nu(U) = \nu(U) \; \sigma(C_{\cA_U}) \; .
\ee

\section{Quadratic forms over $\bZ _N$}\label{app: quadratic}
We want to classify all the quadratic forms $q:\bZ _N^g\rightarrow \bZ _N$ for $N$ prime, namely  symmetric $g\times g$ matrices $\cT$ with $\bZ _N$ entries, up to the equivalence relation
\begin{equation}
    \cT \sim R^T\cT R \ , \ \ \ \ \ R\in \GL(g,\bZ _N) \ . 
\end{equation}
In the case $g=1$ it is easy to see that there are three classes: 0, the perfect squares, and the non perfect squares. Consider now $g=2$. First we put $\cT$ in diagonal form with eigenvalues $p_1,p_2$, which can be swapped by a congruence transformation. If $\cT$ is non singular there are in principle three cases: $p_1,p_2$ are both perfect squares, one of the two is a perfect square but the other is not, and both are not perfect squares. A congruence transformation with $R=\left(\begin{array}{cc}
    a & b \\
    c & d
\end{array}\right)$, $a,b,c,d\neq 0$ preserves the diagonal form if and only if $abp_1+cdp_2=0$, and the new eigenvalues are $p_1'=a^2p_1+c^2p_2$, $p_2'=b^2p_1+d^2p_2$. If $p_1,p_2$ are both perfect squares we can set $p_1=p_2=1$, and to preserve the diagonal form we must have $c=sa$, $b=-sd$ with $s\neq 0$, implying 
\begin{equation}
    p_1'=a^2(s^2+1) \ \ , \ \ \ \ \ \ \ p_2'=d^2(s^2+1) \ .
\end{equation}
We conclude that the case in which $p_1,p_2$ are both perfect squares is equivalent to that in which they are both not perfect square, while the other case form a distinct class. By taking into account the 3 classes at non-maximal rank we get  5 classes.

For generic $g$, if $\cT$ of maximal rank we put it in diagonal form with eigenvalues $p_1,...,p_g$. In principle there are $g+1$ cases, depending on the number $k=0,...,g$ of eigenvalues which are perfect squares. However, using the result at $g=2$ the cases $k$ and $k+2$ are equivalent, while $k$ and $k+1$ are distinct. Thus we get 2 \emph{new} classes at rank $g$ which we did not have at rank $g-1$. Thus the number of classes for $g\times g$ matrices is
\begin{equation}
    n_{\text{classes}}(g)=1+2g\ .
\end{equation}

\section{Fusion Tables}\label{app: Tables}
In this appendix we collect generic results for the fusion algebras of duality defects. We only study the case $g=2$, $N=3$ for definiteness.
A generic fusion between non-invertible defects takes the form
\begin{equation}
    \fD_\cL^{M_1} \times \fD_\cL^{M_2} = \cN ^{r, \pm} \; \cC^\cA  \; \fD_\cL^{M_{1,2}}.
\end{equation}
In the tables below the columns corresponding to the condensate $\cC$ show the generators of the condensed subgroup $\cA$. The numbers $r_i$ are the ranks of the respective defects while $r_c$ is the rank of the condensate.

\paragraph{$\bZ_{4g+2}$ symmetry.}
\begin{equation*}
         \boxed{M_\rho\times M_\rho}
\end{equation*}
\footnotesize
\begin{center}
\begin{longtable}{|c|c|c|c|c|c|c|| c|c|c|c|c|c|c|c|}
 \hline
 $\cL$ & $\text{TQFT}$ & $\cC$ & $r_1$ & $r_2$ & $r_{12}$ & $r_\cC$ &  $\cL$ & $\text{TQFT}$ & $\cC$ & $r_1$ & $r_2$ & $r_{12}$ & $r_{\cC}$\\
 \hline
 \endfirsthead
 \hline
 $\cL$ & $\text{TQFT}$ & $\cC$ & $r_1$ & $r_2$ & $r_{12}$ & $r_\cC$ &  $\cL$ & $\text{TQFT}$ & $\cC$ & $r_1$ & $r_2$ & $r_{12}$ & $r_\cC$\\
 \hline
 \endhead

 \hline
 \endfoot

 \hline
 \endlastfoot
$\left(
\begin{array}{cc}
 0 & 0 \\
 0 & 0 \\
 1 & 0 \\
 0 & 1 \\
\end{array}
\right)$ & $\mathcal{N}^{1,+}$ & $\left(
\begin{array}{c}
 0 \\
 1 \\
\end{array}
\right)$ & $2$ & $2$ & $1$ & $1$ &
$\left(
\begin{array}{cc}
 0 & 0 \\
 0 & 1 \\
 1 & 0 \\
 0 & 1 \\
\end{array}
\right)$ & $\mathcal{N}^{2,-}$ & $\text{Trivial}$ & $2$ & $2$ & $2$ & $0$ \\ \hline
$\left(
\begin{array}{cc}
 0 & 0 \\
 0 & 2 \\
 1 & 0 \\
 0 & 1 \\
\end{array}
\right)$ & $\text{Trivial}$ & $\text{Trivial}$ & $1$ & $1$ & $2$ & $0$ &
$\left(
\begin{array}{cc}
 0 & 1 \\
 1 & 0 \\
 1 & 0 \\
 0 & 1 \\
\end{array}
\right)$ & $\mathcal{N}^{2,-}$ & $\text{Trivial}$ & $2$ & $2$ & $2$ & $0$ \\ \hline
$\left(
\begin{array}{cc}
 0 & 1 \\
 1 & 1 \\
 1 & 0 \\
 0 & 1 \\
\end{array}
\right)$ & $\mathcal{N}^{2,-}$ & $\text{Trivial}$ & $2$ & $2$ & $2$ & $0$ &
$\left(
\begin{array}{cc}
 0 & 1 \\
 1 & 2 \\
 1 & 0 \\
 0 & 1 \\
\end{array}
\right)$ & $\mathcal{N}^{2,-}$ & $\text{Trivial}$ & $2$ & $2$ & $2$ & $0$ \\ \hline
$\left(
\begin{array}{cc}
 0 & 2 \\
 2 & 0 \\
 1 & 0 \\
 0 & 1 \\
\end{array}
\right)$ & $\text{Trivial}$ & $\text{Trivial}$ & $1$ & $1$ & $2$ & $0$ &
$\left(
\begin{array}{cc}
 0 & 2 \\
 2 & 1 \\
 1 & 0 \\
 0 & 1 \\
\end{array}
\right)$ & $\mathcal{N}^{2,+}$ & $\text{Trivial}$ & $2$ & $2$ & $2$ & $0$ \\ \hline
$\left(
\begin{array}{cc}
 0 & 2 \\
 2 & 2 \\
 1 & 0 \\
 0 & 1 \\
\end{array}
\right)$ & $\text{Trivial}$ & $\text{Trivial}$ & $1$ & $1$ & $2$ & $0$ &
$\left(
\begin{array}{cc}
 1 & 0 \\
 0 & 0 \\
 1 & 0 \\
 0 & 1 \\
\end{array}
\right)$ & $\mathcal{N}^{1,+}$ & $\left(
\begin{array}{c}
 1 \\
 1 \\
\end{array}
\right)$ & $2$ & $2$ & $1$ & $1$ \\ \hline
$\left(
\begin{array}{cc}
 1 & 0 \\
 0 & 1 \\
 1 & 0 \\
 0 & 1 \\
\end{array}
\right)$ & $\mathcal{N}^{2,-}$ & $\text{Trivial}$ & $2$ & $2$ & $2$ & $0$ &
$\left(
\begin{array}{cc}
 1 & 0 \\
 0 & 2 \\
 1 & 0 \\
 0 & 1 \\
\end{array}
\right)$ & $\mathcal{N}^{2,-}$ & $\text{Trivial}$ & $2$ & $2$ & $2$ & $0$ \\ \hline
$\left(
\begin{array}{cc}
 1 & 1 \\
 1 & 0 \\
 1 & 0 \\
 0 & 1 \\
\end{array}
\right)$ & $\mathcal{N}^{1,-}$ & $\left(
\begin{array}{c}
 1 \\
 0 \\
\end{array}
\right)$ & $2$ & $2$ & $1$ & $1$ &
$\left(
\begin{array}{cc}
 1 & 1 \\
 1 & 1 \\
 1 & 0 \\
 0 & 1 \\
\end{array}
\right)$ & $\text{Trivial}$ & $\text{Trivial}$ & $1$ & $1$ & $2$ & $0$ \\ \hline
$\left(
\begin{array}{cc}
 1 & 1 \\
 1 & 2 \\
 1 & 0 \\
 0 & 1 \\
\end{array}
\right)$ & $\mathcal{N}^{1,-}$ & $\left(
\begin{array}{c}
 2 \\
 1 \\
\end{array}
\right)$ & $2$ & $2$ & $1$ & $1$ &
$\left(
\begin{array}{cc}
 1 & 2 \\
 2 & 0 \\
 1 & 0 \\
 0 & 1 \\
\end{array}
\right)$ & $\mathcal{N}^{1,-}$ & $\left(
\begin{array}{c}
 2 \\
 1 \\
\end{array}
\right)$ & $2$ & $2$ & $1$ & $1$ \\ \hline
$\left(
\begin{array}{cc}
 1 & 2 \\
 2 & 1 \\
 1 & 0 \\
 0 & 1 \\
\end{array}
\right)$ & $\mathcal{N}^{2,-}$ & $\text{Trivial}$ & $2$ & $2$ & $2$ & $0$ &
$\left(
\begin{array}{cc}
 1 & 2 \\
 2 & 2 \\
 1 & 0 \\
 0 & 1 \\
\end{array}
\right)$ & $\mathcal{N}^{2,-}$ & $\text{Trivial}$ & $2$ & $2$ & $2$ & $0$ \\ \hline
$\left(
\begin{array}{cc}
 2 & 0 \\
 0 & 0 \\
 1 & 0 \\
 0 & 1 \\
\end{array}
\right)$ & $\text{Trivial}$ & $\text{Trivial}$ & $1$ & $1$ & $2$ & $0$ &
$\left(
\begin{array}{cc}
 2 & 0 \\
 0 & 1 \\
 1 & 0 \\
 0 & 1 \\
\end{array}
\right)$ & $\mathcal{N}^{1,-}$ & $\left(
\begin{array}{c}
 1 \\
 1 \\
\end{array}
\right)$ & $2$ & $2$ & $1$ & $1$ \\ \hline
$\left(
\begin{array}{cc}
 2 & 0 \\
 0 & 2 \\
 1 & 0 \\
 0 & 1 \\
\end{array}
\right)$ & $\mathcal{N}^{2,+}$ & $\text{Trivial}$ & $2$ & $2$ & $2$ & $0$ &
$\left(
\begin{array}{cc}
 2 & 1 \\
 1 & 0 \\
 1 & 0 \\
 0 & 1 \\
\end{array}
\right)$ & $\mathcal{N}^{2,-}$ & $\text{Trivial}$ & $2$ & $2$ & $2$ & $0$ \\ \hline
$\left(
\begin{array}{cc}
 2 & 1 \\
 1 & 1 \\
 1 & 0 \\
 0 & 1 \\
\end{array}
\right)$ & $\mathcal{N}^{2,+}$ & $\text{Trivial}$ & $2$ & $2$ & $2$ & $0$ &
$\left(
\begin{array}{cc}
 2 & 1 \\
 1 & 2 \\
 1 & 0 \\
 0 & 1 \\
\end{array}
\right)$ & $\mathcal{N}^{2,-}$ & $\text{Trivial}$ & $2$ & $2$ & $2$ & $0$ \\ \hline
$\left(
\begin{array}{cc}
 2 & 2 \\
 2 & 0 \\
 1 & 0 \\
 0 & 1 \\
\end{array}
\right)$ & $\mathcal{N}^{1,-}$ & $\left(
\begin{array}{c}
 0 \\
 1 \\
\end{array}
\right)$ & $2$ & $2$ & $1$ & $1$ &
$\left(
\begin{array}{cc}
 2 & 2 \\
 2 & 1 \\
 1 & 0 \\
 0 & 1 \\
\end{array}
\right)$ & $\text{Trivial}$ & $\text{Trivial}$ & $1$ & $1$ & $2$ & $0$ \\ \hline
$\left(
\begin{array}{cc}
 2 & 2 \\
 2 & 2 \\
 1 & 0 \\
 0 & 1 \\
\end{array}
\right)$ & $\mathcal{N}^{2,+}$ & $\text{Trivial}$ & $2$ & $2$ & $2$ & $0$ &
$\left(
\begin{array}{cc}
 0 & 0 \\
 1 & 0 \\
 0 & 1 \\
 0 & 0 \\
\end{array}
\right)$ & $\mathcal{N}^{1,+}$ & $\left(
\begin{array}{c}
 1 \\
 0 \\
\end{array}
\right)$ & $2$ & $2$ & $1$ & $1$ \\ \hline
$\left(
\begin{array}{cc}
 0 & 1 \\
 1 & 0 \\
 0 & 1 \\
 0 & 0 \\
\end{array}
\right)$ & $\mathcal{N}^{2,-}$ & $\text{Trivial}$ & $2$ & $2$ & $2$ & $0$ &
$\left(
\begin{array}{cc}
 0 & 2 \\
 1 & 0 \\
 0 & 1 \\
 0 & 0 \\
\end{array}
\right)$ & $\text{Trivial}$ & $\text{Trivial}$ & $1$ & $1$ & $2$ & $0$ \\ \hline
$\left(
\begin{array}{cc}
 2 & 0 \\
 1 & 0 \\
 0 & 1 \\
 0 & 1 \\
\end{array}
\right)$ & $\mathcal{N}^{2,-}$ & $\text{Trivial}$ & $2$ & $2$ & $2$ & $0$ &
$\left(
\begin{array}{cc}
 2 & 1 \\
 1 & 0 \\
 0 & 1 \\
 0 & 1 \\
\end{array}
\right)$ & $\mathcal{N}^{1,+}$ & $\left(
\begin{array}{c}
 2 \\
 1 \\
\end{array}
\right)$ & $2$ & $2$ & $1$ & $1$ \\ \hline
$\left(
\begin{array}{cc}
 2 & 2 \\
 1 & 0 \\
 0 & 1 \\
 0 & 1 \\
\end{array}
\right)$ & $\text{Trivial}$ & $\text{Trivial}$ & $1$ & $1$ & $2$ & $0$ &
$\left(
\begin{array}{cc}
 1 & 0 \\
 1 & 0 \\
 0 & 1 \\
 0 & 2 \\
\end{array}
\right)$ & $\text{Trivial}$ & $\text{Trivial}$ & $1$ & $1$ & $2$ & $0$ \\ \hline
$\left(
\begin{array}{cc}
 1 & 1 \\
 1 & 0 \\
 0 & 1 \\
 0 & 2 \\
\end{array}
\right)$ & $\text{Trivial}$ & $\text{Trivial}$ & $1$ & $1$ & $2$ & $0$ &
$\left(
\begin{array}{cc}
 1 & 2 \\
 1 & 0 \\
 0 & 1 \\
 0 & 2 \\
\end{array}
\right)$ & $\mathcal{N}^{2,+}$ & $\text{Trivial}$ & $2$ & $2$ & $2$ & $0$ \\ \hline
$\left(
\begin{array}{cc}
 1 & 0 \\
 0 & 0 \\
 0 & 0 \\
 0 & 1 \\
\end{array}
\right)$ & $\mathcal{N}^{2,-}$ & $\text{Trivial}$ & $2$ & $2$ & $2$ & $0$ &
$\left(
\begin{array}{cc}
 1 & 0 \\
 0 & 1 \\
 0 & 0 \\
 0 & 1 \\
\end{array}
\right)$ & $\mathcal{N}^{2,-}$ & $\text{Trivial}$ & $2$ & $2$ & $2$ & $0$ \\ \hline
$\left(
\begin{array}{cc}
 1 & 0 \\
 0 & 2 \\
 0 & 0 \\
 0 & 1 \\
\end{array}
\right)$ & $\mathcal{N}^{2,-}$ & $\text{Trivial}$ & $2$ & $2$ & $2$ & $0$ &
$\left(
\begin{array}{cc}
 1 & 0 \\
 0 & 1 \\
 0 & 0 \\
 0 & 0 \\
\end{array}
\right)$ & $\mathcal{N}^{1,+}$ & $\left(
\begin{array}{c}
 0 \\
 1 \\
\end{array}
\right)$ & $2$ & $2$ & $1$ & $1$ \\ \hline
\end{longtable}
\end{center}
\normalsize

\newpage

\paragraph{$D_{12}$ and $(\bZ_2 \times \bZ_6) \ltimes \bZ_2$}
We analyze the noncommutative product (only cases in which the categorical structure fails to commute are shown) and the emergent symmetry $\sigma$ at $\lambda=i$.

\begin{equation*}
         \boxed{M_r\times M_\sigma \ \ \ \ \ M_\sigma \times M_r}
\end{equation*}

\footnotesize
\begin{center}
\begin{longtable}{|c|c|c|c|c|c|c|c|c|}
\hline
$\cL$ & $\text{TQFT}_{1,2}$ & $\cC_{1,2}$ & $\text{TQFT}_{2,1}$ & $\cC_{2,1}$ & $r_1$ & $r_2$ & $\mat{ r_{1,2} \\ r_{2,1} }$ & $\mat{ r_{\cC_{1,2}} \\ r_{\cC_{2,1}}}$  \\

\hline
 \endfirsthead
 \hline
 
 $\cL$ & $\text{TQFT}_{1,2}$ & $\cC_{1,2}$ & $\text{TQFT}_{2,1}$ & $\cC_{2,1}$ & $r_1$ & $r_2$ & $\mat{ r_{1,2} \\ r_{2,1} }$ & $\mat{ r_{\cC_{1,2}} \\ r_{\cC_{2,1}}}$  \\
 
 \hline
 \endhead

 \hline
 \endfoot

 \hline
 \endlastfoot 
$\left(
\begin{array}{cc}
 0 & 0 \\
 0 & 1 \\
 1 & 0 \\
 0 & 1 \\
\end{array}
\right)$ & $\mathcal{N}^{2,+}$ & $\text{Trivial}$ & $\mathcal{N}^{2,-}$ & $\text{Trivial}$ & $2$ & $2$ & $\left(
\begin{array}{c}
 2 \\
 2 \\
\end{array}
\right)$ & $\left(
\begin{array}{c}
 0 \\
 0 \\
\end{array}
\right)$ \\ \hline
$\left(
\begin{array}{cc}
 0 & 0 \\
 0 & 2 \\
 1 & 0 \\
 0 & 1 \\
\end{array}
\right)$ & $\mathcal{N}^{2,+}$ & $\text{Trivial}$ & $\mathcal{N}^{2,-}$ & $\text{Trivial}$ & $2$ & $2$ & $\left(
\begin{array}{c}
 2 \\
 2 \\
\end{array}
\right)$ & $\left(
\begin{array}{c}
 0 \\
 0 \\
\end{array}
\right)$ \\ \hline
$\left(
\begin{array}{cc}
 0 & 1 \\
 1 & 1 \\
 1 & 0 \\
 0 & 1 \\
\end{array}
\right)$ & $\mathcal{N}^{2,-}$ & $\text{Trivial}$ & $\text{Trivial}$ & $\unit_2$ & $2$ & $2$ & $\left(
\begin{array}{c}
 2 \\
 0 \\
\end{array}
\right)$ & $\left(
\begin{array}{c}
 0 \\
 2 \\
\end{array}
\right)$ \\ \hline
$\left(
\begin{array}{cc}
 0 & 2 \\
 2 & 2 \\
 1 & 0 \\
 0 & 1 \\
\end{array}
\right)$ & $\mathcal{N}^{2,-}$ & $\text{Trivial}$ & $\text{Trivial}$ & $\unit_2$ & $2$ & $2$ & $\left(
\begin{array}{c}
 2 \\
 0 \\
\end{array}
\right)$ & $\left(
\begin{array}{c}
 0 \\
 2 \\
\end{array}
\right)$ \\ \hline
$\left(
\begin{array}{cc}
 1 & 0 \\
 0 & 0 \\
 1 & 0 \\
 0 & 1 \\
\end{array}
\right)$ & $\mathcal{N}^{2,-}$ & $\text{Trivial}$ & $\mathcal{N}^{2,+}$ & $\text{Trivial}$ & $2$ & $2$ & $\left(
\begin{array}{c}
 2 \\
 2 \\
\end{array}
\right)$ & $\left(
\begin{array}{c}
 0 \\
 0 \\
\end{array}
\right)$ \\ \hline
$\left(
\begin{array}{cc}
 1 & 1 \\
 1 & 0 \\
 1 & 0 \\
 0 & 1 \\
\end{array}
\right)$ & $\text{Trivial}$ & $\unit_2$ & $\mathcal{N}^{2,-}$ & $\text{Trivial}$ & $2$ & $2$ & $\left(
\begin{array}{c}
 0 \\
 2 \\
\end{array}
\right)$ & $\left(
\begin{array}{c}
 2 \\
 0 \\
\end{array}
\right)$ \\ \hline
$\left(
\begin{array}{cc}
 1 & 1 \\
 1 & 2 \\
 1 & 0 \\
 0 & 1 \\
\end{array}
\right)$ & $\mathcal{N}^{2,+}$ & $\text{Trivial}$ & $\mathcal{N}^{2,-}$ & $\text{Trivial}$ & $2$ & $2$ & $\left(
\begin{array}{c}
 2 \\
 2 \\
\end{array}
\right)$ & $\left(
\begin{array}{c}
 0 \\
 0 \\
\end{array}
\right)$ \\ \hline
$\left(
\begin{array}{cc}
 1 & 2 \\
 2 & 2 \\
 1 & 0 \\
 0 & 1 \\
\end{array}
\right)$ & $\mathcal{N}^{2,-}$ & $\text{Trivial}$ & $\mathcal{N}^{2,+}$ & $\text{Trivial}$ & $2$ & $2$ & $\left(
\begin{array}{c}
 2 \\
 2 \\
\end{array}
\right)$ & $\left(
\begin{array}{c}
 0 \\
 0 \\
\end{array}
\right)$ \\ \hline
$\left(
\begin{array}{cc}
 2 & 0 \\
 0 & 0 \\
 1 & 0 \\
 0 & 1 \\
\end{array}
\right)$ & $\mathcal{N}^{2,-}$ & $\text{Trivial}$ & $\mathcal{N}^{2,+}$ & $\text{Trivial}$ & $2$ & $2$ & $\left(
\begin{array}{c}
 2 \\
 2 \\
\end{array}
\right)$ & $\left(
\begin{array}{c}
 0 \\
 0 \\
\end{array}
\right)$ \\ \hline
$\left(
\begin{array}{cc}
 2 & 1 \\
 1 & 1 \\
 1 & 0 \\
 0 & 1 \\
\end{array}
\right)$ & $\mathcal{N}^{2,-}$ & $\text{Trivial}$ & $\mathcal{N}^{2,+}$ & $\text{Trivial}$ & $2$ & $2$ & $\left(
\begin{array}{c}
 2 \\
 2 \\
\end{array}
\right)$ & $\left(
\begin{array}{c}
 0 \\
 0 \\
\end{array}
\right)$ \\ \hline
$\left(
\begin{array}{cc}
 2 & 2 \\
 2 & 0 \\
 1 & 0 \\
 0 & 1 \\
\end{array}
\right)$ & $\text{Trivial}$ & $\unit_2$ & $\mathcal{N}^{2,-}$ & $\text{Trivial}$ & $2$ & $2$ & $\left(
\begin{array}{c}
 0 \\
 2 \\
\end{array}
\right)$ & $\left(
\begin{array}{c}
 2 \\
 0 \\
\end{array}
\right)$ \\ \hline
$\left(
\begin{array}{cc}
 2 & 2 \\
 2 & 1 \\
 1 & 0 \\
 0 & 1 \\
\end{array}
\right)$ & $\mathcal{N}^{2,+}$ & $\text{Trivial}$ & $\mathcal{N}^{2,-}$ & $\text{Trivial}$ & $2$ & $2$ & $\left(
\begin{array}{c}
 2 \\
 2 \\
\end{array}
\right)$ & $\left(
\begin{array}{c}
 0 \\
 0 \\
\end{array}
\right)$ \\ \hline
$\left(
\begin{array}{cc}
 0 & 0 \\
 1 & 0 \\
 0 & 1 \\
 0 & 0 \\
\end{array}
\right)$ & $\text{Trivial}$ & $\unit_2$ & $\mathcal{N}^{2,-}$ & $\text{Trivial}$ & $2$ & $2$ & $\left(
\begin{array}{c}
 0 \\
 2 \\
\end{array}
\right)$ & $\left(
\begin{array}{c}
 2 \\
 0 \\
\end{array}
\right)$ \\ \hline
$\left(
\begin{array}{cc}
 0 & 1 \\
 1 & 0 \\
 0 & 1 \\
 0 & 0 \\
\end{array}
\right)$ & $\mathcal{N}^{2,+}$ & $\text{Trivial}$ & $\mathcal{N}^{2,-}$ & $\text{Trivial}$ & $2$ & $2$ & $\left(
\begin{array}{c}
 2 \\
 2 \\
\end{array}
\right)$ & $\left(
\begin{array}{c}
 0 \\
 0 \\
\end{array}
\right)$ \\ \hline
$\left(
\begin{array}{cc}
 0 & 2 \\
 1 & 0 \\
 0 & 1 \\
 0 & 0 \\
\end{array}
\right)$ & $\mathcal{N}^{2,+}$ & $\text{Trivial}$ & $\mathcal{N}^{2,-}$ & $\text{Trivial}$ & $2$ & $2$ & $\left(
\begin{array}{c}
 2 \\
 2 \\
\end{array}
\right)$ & $\left(
\begin{array}{c}
 0 \\
 0 \\
\end{array}
\right)$ \\ \hline
$\left(
\begin{array}{cc}
 1 & 0 \\
 0 & 0 \\
 0 & 0 \\
 0 & 1 \\
\end{array}
\right)$ & $\mathcal{N}^{2,-}$ & $\text{Trivial}$ & $\text{Trivial}$ & $\unit_2$ & $2$ & $2$ & $\left(
\begin{array}{c}
 2 \\
 0 \\
\end{array}
\right)$ & $\left(
\begin{array}{c}
 0 \\
 2 \\
\end{array}
\right)$ \\ \hline
$\left(
\begin{array}{cc}
 1 & 0 \\
 0 & 1 \\
 0 & 0 \\
 0 & 1 \\
\end{array}
\right)$ & $\mathcal{N}^{2,-}$ & $\text{Trivial}$ & $\mathcal{N}^{2,+}$ & $\text{Trivial}$ & $2$ & $2$ & $\left(
\begin{array}{c}
 2 \\
 2 \\
\end{array}
\right)$ & $\left(
\begin{array}{c}
 0 \\
 0 \\
\end{array}
\right)$ \\ \hline
$\left(
\begin{array}{cc}
 1 & 0 \\
 0 & 2 \\
 0 & 0 \\
 0 & 1 \\
\end{array}
\right)$ & $\mathcal{N}^{2,-}$ & $\text{Trivial}$ & $\mathcal{N}^{2,+}$ & $\text{Trivial}$ & $2$ & $2$ & $\left(
\begin{array}{c}
 2 \\
 2 \\
\end{array}
\right)$ & $\left(
\begin{array}{c}
 0 \\
 0 \\
\end{array}
\right)$ \\
\end{longtable}
\end{center}
\normalsize
\newpage

\begin{equation*}
    \boxed{M_\sigma \times M_\sigma}
\end{equation*}

\footnotesize
\begin{center}
\begin{longtable}{|c|c|c|c|c|c|c|| c|c|c|c|c|c|c|c|}
\hline
 $\cL$ & $\text{TQFT}$ & $\cC$ & $r_1$ & $r_2$ & $r_{12}$ & $r_{\cC}$ &  $\cL$ & $\text{TQFT}$ & $\cC$ & $r_1$ & $r_2$ & $r_{12}$ & $r_{\cC}$\\
 \hline
 \endfirsthead
 \hline
 $\cL$ & $\text{TQFT}$ & $\cC$ & $r_1$ & $r_2$ & $r_{12}$ & $r_{\cC}$ &  $\cL$ & $\text{TQFT}$ & $\cC$ & $r_1$ & $r_2$ & $r_{12}$ & $r_{\cC}$\\
 \hline
 \endhead
 \hline
 \endfoot
 \hline
 \endlastfoot
 
$\left(
\begin{array}{cc}
 0 & 0 \\
 0 & 0 \\
 1 & 0 \\
 0 & 1 \\
\end{array}
\right)$ & $\text{Trivial}$ & $\unit_2$ & $2$ & $2$ & $0$ & $2$ & $\left(
\begin{array}{cc}
 2 & 0 \\
 0 & 2 \\
 1 & 0 \\
 0 & 1 \\
\end{array}
\right)$ & $\mathcal{N}^{2,-}$ & $\text{Trivial}$ & $2$ & $2$ & $2$ & $0$ \\ \hline
$\left(
\begin{array}{cc}
 0 & 0 \\
 0 & 1 \\
 1 & 0 \\
 0 & 1 \\
\end{array}
\right)$ & $\mathcal{N}^{2,-}$ & $\text{Trivial}$ & $2$ & $2$ & $2$ & $0$ & $\left(
\begin{array}{cc}
 2 & 1 \\
 1 & 0 \\
 1 & 0 \\
 0 & 1 \\
\end{array}
\right)$ & $\mathcal{N}^{1,+}$ & $\text{Trivial}$ & $1$ & $1$ & $1$ & $0$ \\ \hline
$\left(
\begin{array}{cc}
 0 & 0 \\
 0 & 2 \\
 1 & 0 \\
 0 & 1 \\
\end{array}
\right)$ & $\mathcal{N}^{2,-}$ & $\text{Trivial}$ & $2$ & $2$ & $2$ & $0$ & $\left(
\begin{array}{cc}
 2 & 1 \\
 1 & 1 \\
 1 & 0 \\
 0 & 1 \\
\end{array}
\right)$ & $\mathcal{N}^{2,-}$ & $\text{Trivial}$ & $2$ & $2$ & $2$ & $0$ \\ \hline
$\left(
\begin{array}{cc}
 0 & 1 \\
 1 & 0 \\
 1 & 0 \\
 0 & 1 \\
\end{array}
\right)$ & $\mathcal{N}^{2,-}$ & $\text{Trivial}$ & $2$ & $2$ & $2$ & $0$ & $\left(
\begin{array}{cc}
 2 & 1 \\
 1 & 2 \\
 1 & 0 \\
 0 & 1 \\
\end{array}
\right)$ & $\mathcal{N}^{2,-}$ & $\text{Trivial}$ & $2$ & $2$ & $2$ & $0$ \\ \hline
$\left(
\begin{array}{cc}
 0 & 1 \\
 1 & 1 \\
 1 & 0 \\
 0 & 1 \\
\end{array}
\right)$ & $\mathcal{N}^{2,-}$ & $\text{Trivial}$ & $2$ & $2$ & $2$ & $0$ & $\left(
\begin{array}{cc}
 2 & 2 \\
 2 & 0 \\
 1 & 0 \\
 0 & 1 \\
\end{array}
\right)$ & $\mathcal{N}^{2,-}$ & $\text{Trivial}$ & $2$ & $2$ & $2$ & $0$ \\ \hline
$\left(
\begin{array}{cc}
 0 & 1 \\
 1 & 2 \\
 1 & 0 \\
 0 & 1 \\
\end{array}
\right)$ & $\mathcal{N}^{1,-}$ & $\text{Trivial}$ & $1$ & $1$ & $1$ & $0$ & $\left(
\begin{array}{cc}
 2 & 2 \\
 2 & 1 \\
 1 & 0 \\
 0 & 1 \\
\end{array}
\right)$ & $\mathcal{N}^{2,-}$ & $\text{Trivial}$ & $2$ & $2$ & $2$ & $0$ \\ \hline
$\left(
\begin{array}{cc}
 0 & 2 \\
 2 & 0 \\
 1 & 0 \\
 0 & 1 \\
\end{array}
\right)$ & $\mathcal{N}^{2,-}$ & $\text{Trivial}$ & $2$ & $2$ & $2$ & $0$ & $\left(
\begin{array}{cc}
 2 & 2 \\
 2 & 2 \\
 1 & 0 \\
 0 & 1 \\
\end{array}
\right)$ & $\text{Trivial}$ & $\unit_2$ & $2$ & $2$ & $0$ & $2$ \\ \hline
$\left(
\begin{array}{cc}
 0 & 2 \\
 2 & 1 \\
 1 & 0 \\
 0 & 1 \\
\end{array}
\right)$ & $\mathcal{N}^{1,+}$ & $\text{Trivial}$ & $1$ & $1$ & $1$ & $0$ & $\left(
\begin{array}{cc}
 0 & 0 \\
 1 & 0 \\
 0 & 1 \\
 0 & 0 \\
\end{array}
\right)$ & $\mathcal{N}^{2,-}$ & $\text{Trivial}$ & $2$ & $2$ & $2$ & $0$ \\ \hline
$\left(
\begin{array}{cc}
 0 & 2 \\
 2 & 2 \\
 1 & 0 \\
 0 & 1 \\
\end{array}
\right)$ & $\mathcal{N}^{2,-}$ & $\text{Trivial}$ & $2$ & $2$ & $2$ & $0$ & $\left(
\begin{array}{cc}
 0 & 1 \\
 1 & 0 \\
 0 & 1 \\
 0 & 0 \\
\end{array}
\right)$ & $\mathcal{N}^{2,-}$ & $\text{Trivial}$ & $2$ & $2$ & $2$ & $0$ \\ \hline
$\left(
\begin{array}{cc}
 1 & 0 \\
 0 & 0 \\
 1 & 0 \\
 0 & 1 \\
\end{array}
\right)$ & $\mathcal{N}^{2,-}$ & $\text{Trivial}$ & $2$ & $2$ & $2$ & $0$ & $\left(
\begin{array}{cc}
 0 & 2 \\
 1 & 0 \\
 0 & 1 \\
 0 & 0 \\
\end{array}
\right)$ & $\mathcal{N}^{2,-}$ & $\text{Trivial}$ & $2$ & $2$ & $2$ & $0$ \\ \hline
$\left(
\begin{array}{cc}
 1 & 0 \\
 0 & 1 \\
 1 & 0 \\
 0 & 1 \\
\end{array}
\right)$ & $\mathcal{N}^{2,-}$ & $\text{Trivial}$ & $2$ & $2$ & $2$ & $0$ & $\left(
\begin{array}{cc}
 2 & 0 \\
 1 & 0 \\
 0 & 1 \\
 0 & 1 \\
\end{array}
\right)$ & $\mathcal{N}^{2,-}$ & $\text{Trivial}$ & $2$ & $2$ & $2$ & $0$ \\ \hline
$\left(
\begin{array}{cc}
 1 & 0 \\
 0 & 2 \\
 1 & 0 \\
 0 & 1 \\
\end{array}
\right)$ & $\mathcal{N}^{1,+}$ & $\text{Trivial}$ & $1$ & $1$ & $1$ & $0$ & $\left(
\begin{array}{cc}
 2 & 1 \\
 1 & 0 \\
 0 & 1 \\
 0 & 1 \\
\end{array}
\right)$ & $\mathcal{N}^{2,-}$ & $\text{Trivial}$ & $2$ & $2$ & $2$ & $0$ \\ \hline
$\left(
\begin{array}{cc}
 1 & 1 \\
 1 & 0 \\
 1 & 0 \\
 0 & 1 \\
\end{array}
\right)$ & $\mathcal{N}^{2,-}$ & $\text{Trivial}$ & $2$ & $2$ & $2$ & $0$ & $\left(
\begin{array}{cc}
 2 & 2 \\
 1 & 0 \\
 0 & 1 \\
 0 & 1 \\
\end{array}
\right)$ & $\mathcal{N}^{2,-}$ & $\text{Trivial}$ & $2$ & $2$ & $2$ & $0$ \\ \hline
$\left(
\begin{array}{cc}
 1 & 1 \\
 1 & 1 \\
 1 & 0 \\
 0 & 1 \\
\end{array}
\right)$ & $\text{Trivial}$ & $\unit_2$ & $2$ & $2$ & $0$ & $2$ & $\left(
\begin{array}{cc}
 1 & 0 \\
 1 & 0 \\
 0 & 1 \\
 0 & 2 \\
\end{array}
\right)$ & $\text{Trivial}$ & $\text{Trivial}$ & $0$ & $0$ & $0$ & $0$ \\ \hline
$\left(
\begin{array}{cc}
 1 & 1 \\
 1 & 2 \\
 1 & 0 \\
 0 & 1 \\
\end{array}
\right)$ & $\mathcal{N}^{2,-}$ & $\text{Trivial}$ & $2$ & $2$ & $2$ & $0$ & $\left(
\begin{array}{cc}
 1 & 1 \\
 1 & 0 \\
 0 & 1 \\
 0 & 2 \\
\end{array}
\right)$ & $\text{Trivial}$ & $\unit_2$ & $2$ & $2$ & $0$ & $2$ \\ \hline
$\left(
\begin{array}{cc}
 1 & 2 \\
 2 & 0 \\
 1 & 0 \\
 0 & 1 \\
\end{array}
\right)$ & $\mathcal{N}^{1,-}$ & $\text{Trivial}$ & $1$ & $1$ & $1$ & $0$ & $\left(
\begin{array}{cc}
 1 & 2 \\
 1 & 0 \\
 0 & 1 \\
 0 & 2 \\
\end{array}
\right)$ & $\text{Trivial}$ & $\unit_2$ & $2$ & $2$ & $0$ & $2$ \\ \hline
$\left(
\begin{array}{cc}
 1 & 2 \\
 2 & 1 \\
 1 & 0 \\
 0 & 1 \\
\end{array}
\right)$ & $\mathcal{N}^{2,-}$ & $\text{Trivial}$ & $2$ & $2$ & $2$ & $0$ & $\left(
\begin{array}{cc}
 1 & 0 \\
 0 & 0 \\
 0 & 0 \\
 0 & 1 \\
\end{array}
\right)$ & $\mathcal{N}^{2,-}$ & $\text{Trivial}$ & $2$ & $2$ & $2$ & $0$ \\ \hline
$\left(
\begin{array}{cc}
 1 & 2 \\
 2 & 2 \\
 1 & 0 \\
 0 & 1 \\
\end{array}
\right)$ & $\mathcal{N}^{2,-}$ & $\text{Trivial}$ & $2$ & $2$ & $2$ & $0$ & $\left(
\begin{array}{cc}
 1 & 0 \\
 0 & 1 \\
 0 & 0 \\
 0 & 1 \\
\end{array}
\right)$ & $\mathcal{N}^{2,-}$ & $\text{Trivial}$ & $2$ & $2$ & $2$ & $0$ \\ \hline
$\left(
\begin{array}{cc}
 2 & 0 \\
 0 & 0 \\
 1 & 0 \\
 0 & 1 \\
\end{array}
\right)$ & $\mathcal{N}^{2,-}$ & $\text{Trivial}$ & $2$ & $2$ & $2$ & $0$ & $\left(
\begin{array}{cc}
 1 & 0 \\
 0 & 2 \\
 0 & 0 \\
 0 & 1 \\
\end{array}
\right)$ & $\mathcal{N}^{2,-}$ & $\text{Trivial}$ & $2$ & $2$ & $2$ & $0$ \\ \hline
$\left(
\begin{array}{cc}
 2 & 0 \\
 0 & 1 \\
 1 & 0 \\
 0 & 1 \\
\end{array}
\right)$ & $\mathcal{N}^{1,-}$ & $\text{Trivial}$ & $1$ & $1$ & $1$ & $0$ & $\left(
\begin{array}{cc}
 1 & 0 \\
 0 & 1 \\
 0 & 0 \\
 0 & 0 \\
\end{array}
\right)$ & $\text{Trivial}$ & $\unit_2$ & $2$ & $2$ & $0$ & $2$ \\ \hline
\end{longtable}
\end{center}
\normalsize

\bibliographystyle{ytphys}
\baselineskip=0.99\baselineskip
\bibliography{TopGravity}

\end{document}